\documentclass[final,5p,times,twocolumn,figuresright]{elsarticle}

\usepackage{amssymb}

\journal{High Energy Density Physics}
\usepackage[english]{babel}
\usepackage[utf8x]{inputenc}
\usepackage[T1]{fontenc}

\usepackage{amsmath}
\usepackage{graphicx}
\usepackage[colorlinks=true, allcolors=blue]{hyperref}
\usepackage{multirow}
\usepackage{rotating}
\usepackage{pdflscape}
\usepackage{rotating}
\usepackage{longtable}
\usepackage{lscape}
\usepackage{xcolor}
\usepackage[normalem]{ulem}

\newcommand{\new}[1]{#1}
\newcommand{\old}[1]{}

\begin{document}



\title{Review of the First Charged-Particle Transport Coefficient Comparison Workshop}


\author[LLNL]{P. E. Grabowski}
\address[LLNL]{Lawrence Livermore National Laboratory, Livermore, CA 94550, USA}
\ead{grabowski5@llnl.gov}
\author[SNL]{S. B. Hansen}
\address[SNL]{Sandia National Laboratory, Albuquerque, NM 87185, USA}
\author[MSU]{M. S. Murillo}
\address[MSU]{Department of Computational Mathematics, Science and Engineering, Michigan State University, East Lansing, Michigan 48824, USA}
\author[SJSU]{L.~G.~Stanton}
\address[SJSU]{Department of Mathematics and Statistics, San Jos\'e State University, San Jos\'e, California 95192, USA}
\author[LLNL]{F. R. Graziani}
\author[LLNL]{A. B. Zylstra}
\address[LANL]{Los Alamos National Laboratory, Los Alamos, NM 87545, USA}
\author[Iowa]{S. D. Baalrud}
\address[Iowa]{Department of Physics and Astronomy, University of Iowa, Iowa City, IA 52242, USA}
\author[CEA]{\new{P. Arnault}}
\address[CEA]{D\'{e}partement de Physique Th\'{e}orique et Appliqu\'{e}e, CEA, DAM, DIF \^{I}le-de-France BP12, 91680 Bruy\'{e}res-le-Ch\^{a}tel Cedex, France}
\author[SNL]{A. D. Baczewski}
\author[LLNL]{L. X. Benedict}
\author[CEA2]{C. Blancard}
\address[CEA2]{CEA, DAM, DIF, F-91297 Arpajon, France}
\author[LANL]{O. \v{C}ert\'{\i}k}
\author[CEA]{J. Cl\'erouin}
\author[LANL]{L. A. Collins}
\author[LLNL]{S. Copeland}
\author[LLNL]{A. A. Correa}
\author[NUDT]{J. Dai}
\address[NUDT]{Department of Physics, National University of Defense Technology, Changsha, Hunan 410073, People's Republic of China}
\author[LANL]{J. Daligault}
\author[SNL]{M. P. Desjarlais} 
\author[NRC]{M. W. C. Dharma-wardana}
\address[NRC]{National Research Council of Canada, Ottawa, Ontario, Canada, K1A 0R6.}
\author[CEA2]{G. Faussurier}
\author[LANL]{J. Haack}
\author[LLNL]{\old{S. Haan}}
\author[LLNL]{T. Haxhimali}
\author[LANL]{A. Hayes-Sterbenz}
\author[NUDT]{Y. Hou}
\author[LLE]{S. X. Hu}
\address[LLE]{Laboratory for Laser Energetics, University of Rochester, 250 East River Road, Rochester, New York 14623, USA}
\author[SNL]{D. Jensen}
\author[LANL]{G. Jungman}
\author[Imp]{G. Kagan}
\address[Imp]{Imperial College, London, SW7 2AZ, United Kingdom}
\author[NUDT]{D. Kang}
\author[LANL]{J. D. Kress}
\author[NUDT]{Q. Ma}
\author[LANL]{M. Marciante}
\author[LANL]{E. Meyer}
\author[LLNL]{R. E. Rudd}
\author[LANL]{D. Saumon}
\author[SNL]{L. Shulenburger}
\author[LANL]{R. L. Singleton Jr.}
\author[LANL]{T. Sjostrom}
\author[MSU]{L. J. Stanek}
\author[LANL]{C. E. Starrett}
\author[LANL]{C. Ticknor}
\author[LANL]{S. Valaitis}
\author[LANL]{J. Venzke}
\author[LANL]{A. White}


\begin{abstract}
We present the results of the first Charged-Particle Transport Coefficient Code Comparison Workshop, which was held in Albuquerque, NM October 4-6, 2016. In this first workshop, scientists from eight institutions and four countries gathered to compare calculations of transport coefficients including thermal and electrical conduction, electron-ion coupling, inter-ion diffusion, ion viscosity, and charged particle stopping powers. Here, we give general background on Coulomb coupling and computational expense, review where 
some transport coefficients appear in hydrodynamic equations, and present the submitted data. Large variations are found when either the relevant Coulomb coupling parameter is large or computational expense causes difficulties. Understanding the general accuracy and uncertainty associated with such transport coefficients is important for quantifying errors in hydrodynamic simulations of inertial confinement fusion and high-energy density experiments.
\end{abstract}





\maketitle


\section{Introduction}

Charged-particle transport is a key part of high-energy-density plasma science. Transport coefficients feed into both radiation-hydrodynamic simulations and diagnostic data interpretation, but neither uncertainties in these coefficients nor the features of generally reliable transport models are well established. 
They are particularly important 
when inter-particle correlations of the plasma are weak 
enough that there are important deviations from the ideal fluid (Euler) limit, but strong enough so that the simplest kinetic
treatments (e.g. the Vlasov equation) are no longer valid.
Reliance on theoretical predictions is driven by the challenges of experimentally isolating various transport processes, together
with the paucity of experimental data of any kind at well-characterized extreme conditions.
These coefficients have particular impact on the field of inertial confinement fusion, feeding into the development of instabilities 
and the overall energy balance of burning fusion plasma.
Crucial processes include thermal and electrical conduction, electron-ion coupling, inter-ion diffusion, ion viscosity, and charged particle stopping. 

The first charged-particle transport coefficient workshop (CPTCW-16)\old{, 
modeled after the successful non-local thermodynamic equilibrium (non-LTE) opacity code comparison workshops, \cite{nlte}}
was established to examine theoretical uncertainties in our predictive ability. \new{Workshops of this sort
have become a tradition in dense plasma physics. Recently, there has been the 2016 kinetics workshop \cite{RinderknechtEtAl2018} and 
2017 equation-of-state (EOS) workshop \cite{GaffneyEtAl2018}, which have built on the successes of 
the long-running non-local thermodynamic equilibrium (non-LTE) opacity code comparison workshops \cite{LeeEtAl1997,BowenEtAl2003,BowenEtAl2006,RubianoEtAl2007,FontesEtAl2009,NLTE6,ChungEtAl2013,nlte,HansenEtAl2020}
and similar WorkOp LTE opacity workshop \cite{Rose1994,Rickert1995,SerdukeEtAl2000}}.
A set of test cases spanning a range of ionization, coupling, and degeneracy regimes was selected with the aim of establishing the present state of agreement among various theoretical approaches. In addition to this goal, the workshop aimed to quantify uncertainties in calculated transport coefficients, address the strengths and limitations of different approaches, provide a forum for discussions, and identify research priorities for inertial confinement fusion (ICF) and high energy-density (HED) science.


\section{Test Cases}

Three plasma compositions were chosen across wide temperature and density variations, including pure hydrogen, pure carbon, and an equimolar carbon-hydrogen mixture. These are summarized in Table  \ref{tab:cases}.
Plasmas across these conditions are relevant to fuel and ablator materials important to ICF. In particular, the low-density, high-temperature hydrogen case approaches the conditions of recent Omega experiments measuring stopping powers \cite{FrenjeEtAl2015, FrenjeEtAl2019} and the low-density, moderate-temperature carbon case is at conditions relevant to recent Omega experiments on heated beryllium \cite{ZylstraEtAl2015,ZylstraEtAl2016}, with additional relevance to thermal conductivity experiments in plastic and beryllium \cite{PingEtAl2015,McKelveyEtAl2017}.

\begin{table}
\centering
\begin{tabular}{c|l|l}
element & density (g/cm$^3$) & temperature (eV)  \\\hline
C & 0.1, 1, 10, 100 & 0.2, 2, 20, 200, 2000  \\ 
H & 0.1, 1, 10, 100 & 0.2, 2, 20, 200, 2000  \\ 
CH & 0.1, 1, 10, 100 & 0.2, 2, 20, 200, 2000  
\end{tabular}
\caption{\label{tab:cases} Summary of workshop cases.}
\end{table}

More than thirty researchers attended the workshop, representing eight institutions; participation is summarized in Table \ref{tab:contributors}. There were averages of $5.5$, $6.3$, $8.6$, $8.6$, and $2.9$ contributions for the $60$ different cases, 
for electrical conductivity, thermal conductivity, viscosity, diffusion, and stopping power, respectively. Models expressed as  analytic formulae, which can be used inline in hydrodynamic codes, to those employing a range of potentials in molecular dynamics were represented. While several approaches presented data for all cases, most contributions only provided ionic transport (diffusion,viscosity) or electronic transport (thermal and electrical conductivities) or stopping powers.  \new{The results reported here represent output from the contributed codes at the time of the workshop; any subsequently published updates to those codes are noted in the table of contributors but not represented in the present paper.}

\begin{table*}[t]
  \centering\begin{tabular}{l|l|l}
Contributors & Institution & Description \\\hline
Baalrud & U Iowa & Effective Potential Theory (EPT) with Average Atom potentials \cite{BaalrudDaligault2013,DaligaultEtAl2016}\\
 Baczewski, Jensen,  & \multirow{2}{*}{SNL} & \multirow{2}{*}{Ehrenfest-TDDFT (VASP-TDDFT ) (stopping powers) \cite{MagyarEtAl2016,BaczewskiEtAl2016} }\\
\quad Shulenburger &&\\
\multirow{2}{*}{Cl\'erouin\new{, Arnault}} & \multirow{2}{*}{CEA} & Global One-Component Plasma (PIJ) \cite{Arnault2013}, \\&&\quad orbital-free Thomas Fermi \cite{LAMB07,LAMB13} \\ 
Copeland & LLNL & various analytic models \cite{ChapmanCowling1939,LeeMore1984,PaquetteEtAl1986} \\
Copeland, Stanton,  & LLNL, SJSU & \multirow{2}{*}{Effective Yukawa T-Matrix \cite{StantonMurillo2016}} \\
\quad Murillo, Stanek & \quad MSU&\\
Daligault & LANL & classical molecular dynamics (MD) with Average Atom potentials \\ 
 Desjarlais & SNL & SNL-modified quantum MD with Kubo-Greenwood  (VASP-SNL) \cite{DesjarlaisEtAl2017}\\ 
Dharma-wardana & NRC Canada & Neutral Pseudo-Atom \cite{Perrot1993,PerrotDharmawardana1995,PerrotDharmawardana1999} \\ 
Faussurier, Blancard & CEA & Two-component electron-ion Average-Atom (SCAALP) \cite{BlancardFaussurier2004,FaussurierEtAl2010} \\ 
Grabowski, Starrett, Saumon & LLNL\new{, LANL} & \old{Psuedo}\new{Pseudo}-Atom MD density with strong scattering corrections \cite{ZylstraEtAl2015} \\ 
\multirow{2}{*}{Hansen} & \multirow{2}{*}{SNL} & Average Atom and Neutral Pseudo-Atom \\&&\quad with Ziman conductivity (\old{Muze and Bemuze} \new{MuZe and BeMuZe})\\ 
Haxhimali, Rudd & LLNL & Hybrid Kinetic Molecular Dynamics (KMD) \cite{HaxhimaliEtAl2015} \\ 
Hayes, Singleton, Jungman & LANL & degenerate Brown-Preston-Singleton (BPS) (Stopping) \cite{BrownEtAl2005} \\ 
Hou & NUDT & average atom hypernetted chain \cite{HouEtAl2015}\\ 
\multirow{2}{*}{Hu} & \multirow{2}{*}{LLE} & Spitzer-Lee-More and quantum MD\\
&&\quad with Kubo-Greenwood (VASP) \cite{HuEtAl2014,HuEtAl2016,DingEtAl2018,WhiteEtAl2018} \\ 
\multirow{2}{*}{Kang, Dai} & \multirow{2}{*}{NUDT} & Path Integral Molecular Dynamics (PIMD), \\
&&\quad Quantum MD (Quantum Espresso) \cite{KangDai2018,MaEtAl2018}\\ 
Ma, Dai & NUDT & electron Force Field (eFF) \cite{MaEtAl2019}\\ 
Marciante & LANL & Thomas-Fermi-Yukawa MD \\ 
Meyer, Collins & LANL & quantum MD with Kubo-Greenwood (VASP) \cite{HuEtAl2014,HuEtAl2016,DingEtAl2018,WhiteEtAl2018} \\ 
\multirow{3}{*}{Sjostrom} & \multirow{3}{*}{LANL} & Quantum MD (Quantum Espresso),\\ &&\quad Kohn-Sham MD with nonlocal corrections \cite{SjostromDaligault2015}, \\&&\quad Thomas-Fermi MD \\ 
Starrett, Saumon & LANL & Pseudo-Atom MD \cite{StarrettEtAl2015, StarrettSaumon2016,Starrett2016}\new{; updated \cite{Starrett2017}} \\ 
Ticknor, \v{C}ert\'{\i}k, White & \multirow{2}{*}{LANL} & \multirow{2}{*}{Orbital-free molecular MD with Thomas-Fermi-Dirac functional \cite{LambertEtAl2006,LambertRecoules2012,MeyerEtAl2014,TicknorEtAl2016} }\\
\quad Collins, Venzke, Valaitis&&
\end{tabular}
\caption{\label{tab:contributors} Summary of contributors and models.}
\end{table*}


\section{\label{SecDimensionless}Dimensionless Parameters and Test Case Coverage in Plasma Parameter Space}

HED plasmas span many orders of magnitude in density and temperature, 
including regimes for which different approximations are valid. 
In this section we will discuss dimensionless parameters that capture this diversity and show where our test cases fall.

We can characterize the importance of interactions with the Coulomb coupling parameter
\begin{equation}
\Gamma_{jk}=\left|\frac{q_jq_k}{a_{jk} T_{jk,eff}}\right|,
\end{equation}
which gives the relative magnitude of the potential energy of neighboring particles to their kinetic 
energy. Here, $j$ and $k$ are indices for particle types. For ions, $q_i = \bar{Z}_i e$, while for electrons,
$q_e=e$, where $e$ is the magnitude of the electron charge and $\bar{Z}_i$ is the mean ionization the ion,
which for convenience we set to More's Thomas-Fermi fit \cite{More1981} (such a choice is not unique \cite{MurilloEtAl2013}). The typical separation between neighbors is 
approximated by
\begin{equation}
a_{jk}=\left[\frac{6}{4\pi(n_j+n_k)}\right]^{1/3},
\end{equation}
where for ions, $n_i = \rho_i/m_i$ and for electrons, $n_e = \sum_i Z_i\rho_i/m_i$ with the sum over the ions,
$Z_i$ is the bare ion charge, 
$\rho_i$ is the mass density of ion $i$ and $m_i$ is its mass. 
A non-unique effective temperature, $T_{jk,eff}$, is a measure of the relative kinetic energy of collisions between particles of types
$j$ and $k$. 
We set
\begin{equation}
T_{jk,eff} = \sqrt{T^2+\frac{2}{25} (T_{j,F}^2+T_{k,F}^2)},
\end{equation}
where 
\begin{equation}
k_B T_{j,F} = \frac{\hbar^2}{2m_j}(3\pi^2 n_{j})^{2/3},
\end{equation}
$k_B$ is Boltzmann's constant, $m_j$ is the mass of particles of type $j$, and $n_{j}$ is their number density. This effective temperature is a simple interpolation between the zero and infinite temperature limits.

\begin{figure}
\centering
\includegraphics[width=.95\columnwidth]{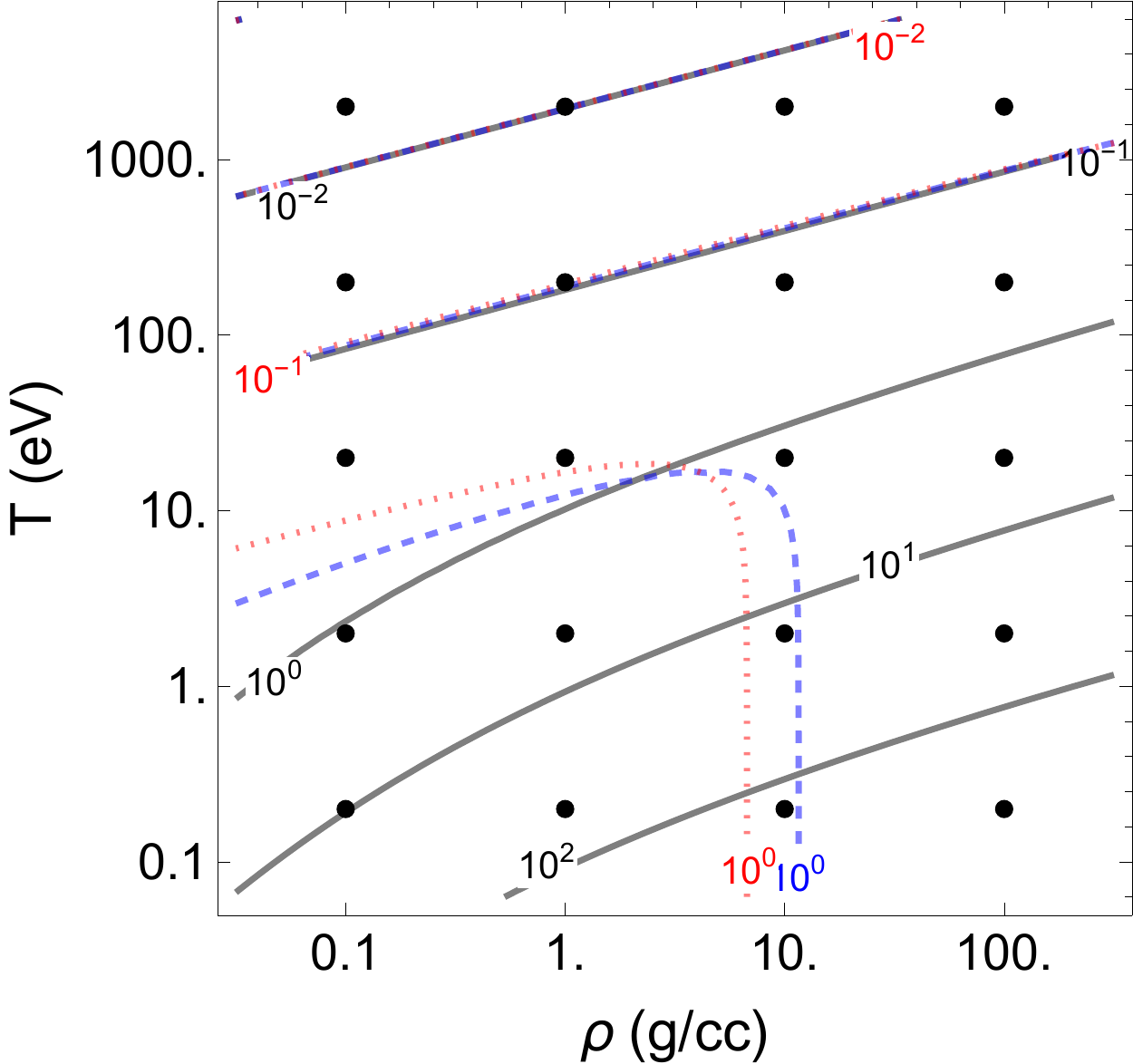}
\caption{\label{HGamma} Ion-Ion (solid black), ion-electron (dashed blue), and 
electron-electron (dotted red) Coulomb coupling parameters for pure hydrogen.
See text for definitions. Large Coulomb coupling parameters indicate where 
analytic expressions break down. 
The workshop cases are at the positions of the black dots. The ion-ion coupling parameter
is large at high densities and low temperatures, whereas the electron-ion and electron-electron
coupling parameters are never much above unity in the high energy density regime due to 
the Pauli exclusion principle.}
\end{figure}

\begin{figure}
\centering
\includegraphics[width=.95\columnwidth]{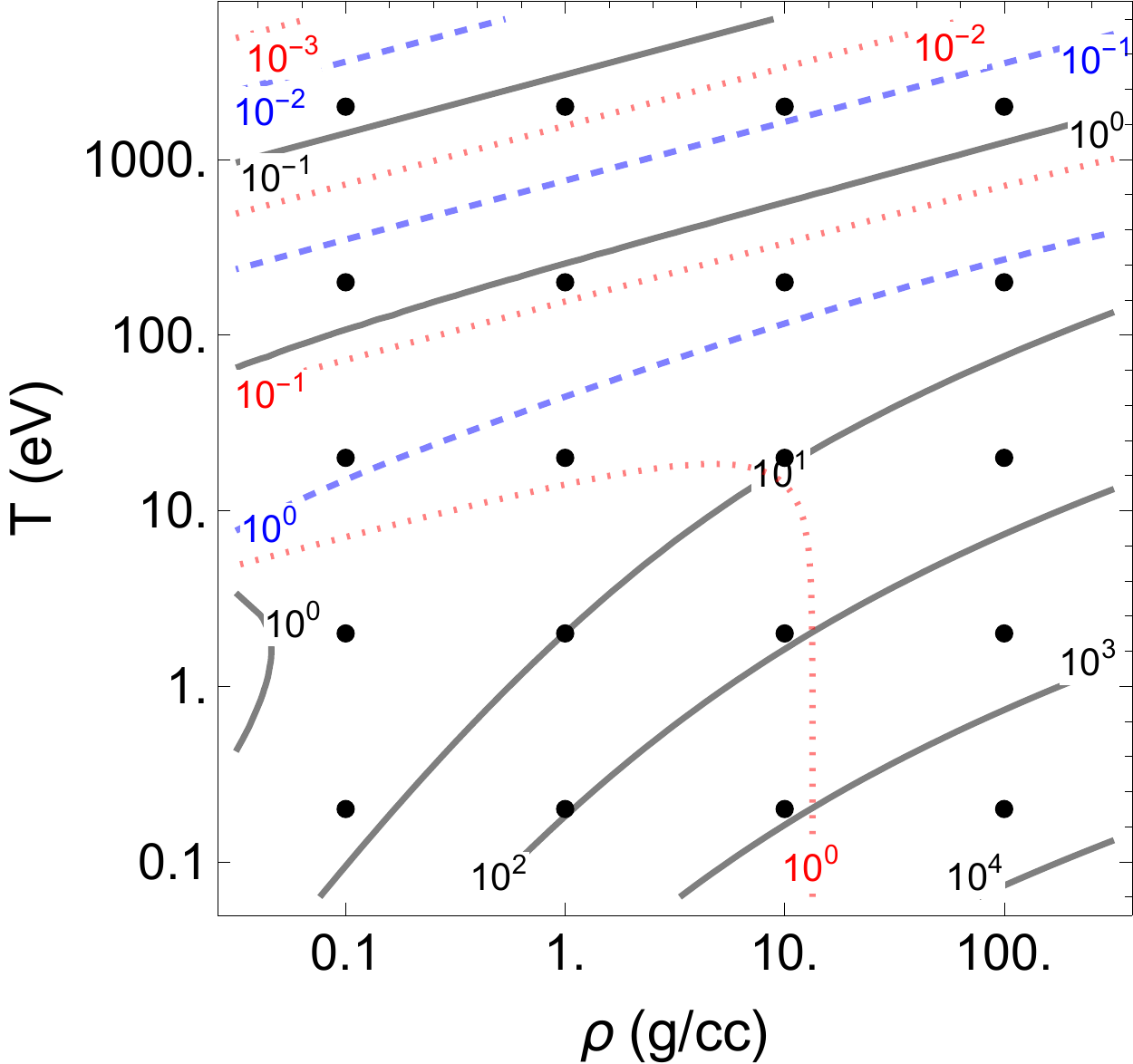}
\caption{\label{CGamma} Same as Fig. \ref{HGamma} but for pure carbon. Note the broad range of parameters for 
which the ion-ion coupling is close to unity, due to roughly equal and opposite effects of temperature and mean
ionization.}
\end{figure}

These coupling parameters fall into three categories: ion-ion, electron-ion, and electron-electron, depending on
the particle types, $j$ and $k$. Different transport coefficients are more sensitive to some of these Coulomb coupling parameters 
than others and so become harder to calculate in different regimes.
As ion viscosity and ion diffusion are mediated mainly by ion-ion collisions, the ion-ion Coulomb coupling 
parameter is most important. Of course, electron screening also plays a role. The main hindrance to 
electrical conductivity is electron-ion scattering, 
and so the electron-ion Coulomb coupling parameter 
matters for this property. Stopping power of fast ions is dominated by projectile-electron collisions, which are similar
to the target ion-electron collisions except for having a different energy scale: the relative kinetic
energy of the ion and the electrons ($K_{rel}=\mu \langle v_{rel}^2 \rangle/2$, where $\mu$ is the reduced mass of the projectile and target particles and $\langle v_{rel}^ 2 \rangle$ is the ensemble mean of the square of their relative velocities), suggesting one replace $T_{jk,eff}$ with $K_{rel}$ in the 
coupling parameter expression. So the ion-electron coupling parameter shown in Figs. \ref{HGamma} and \ref{CGamma} 
should be considered upper bounds for stopping power and mostly relevant to low energy (at and below the Bragg Peak) ions. 
Thermal conductivity is sensitive to both the electron-electron and electron-ion parameters. 

For small $\Gamma$'s ($\ll 1$), it is valid to introduce screening and collision physics through cutoffs in collision integrals,
which leads to a Coulomb logarithm.  As $\Gamma_{jk}$ approaches unity, such treatments break down
and one should use more sophisticated methods, the limits of which are explored in this article. 
The relevant values of $\Gamma_{jk}$ for our study are shown in Figs. \ref{HGamma} and \ref{CGamma} for
pure hydrogen and carbon plasmas, respectively. We do not show the CH mixture case, but it is qualitatively 
similar to the other two. We see that all three (ion-ion, ion-electron, and electron-electron) coupling parameters
are small in the high-temperature low-density limit. 
We note that for carbon and any other \old{multielectron}\new{multi-electron} atom, there is a large plateau of moderate ion-ion coupling
due to the roughly equal and opposite effects of temperature and ionization \cite{ARNA13, ClerouinEtAl2015}.
In the low-temperature high-density limit, the ions are very strongly coupled and crystalize, causing long-range correlations 
well beyond the screening length and caging effects. However, because the electrons are degenerate in this limit and so have 
kinetic energies of order the Fermi energy, they become weakly coupled at large densities. This means that
conductivity and stopping power models which take into account degeneracy can be accurate. 
Electrons are only strongly coupled in the nondegenerate low-temperature low-density limit, which is outside the field of high energy density physics.

\begin{figure}
\centering
\includegraphics[width=.95\columnwidth]{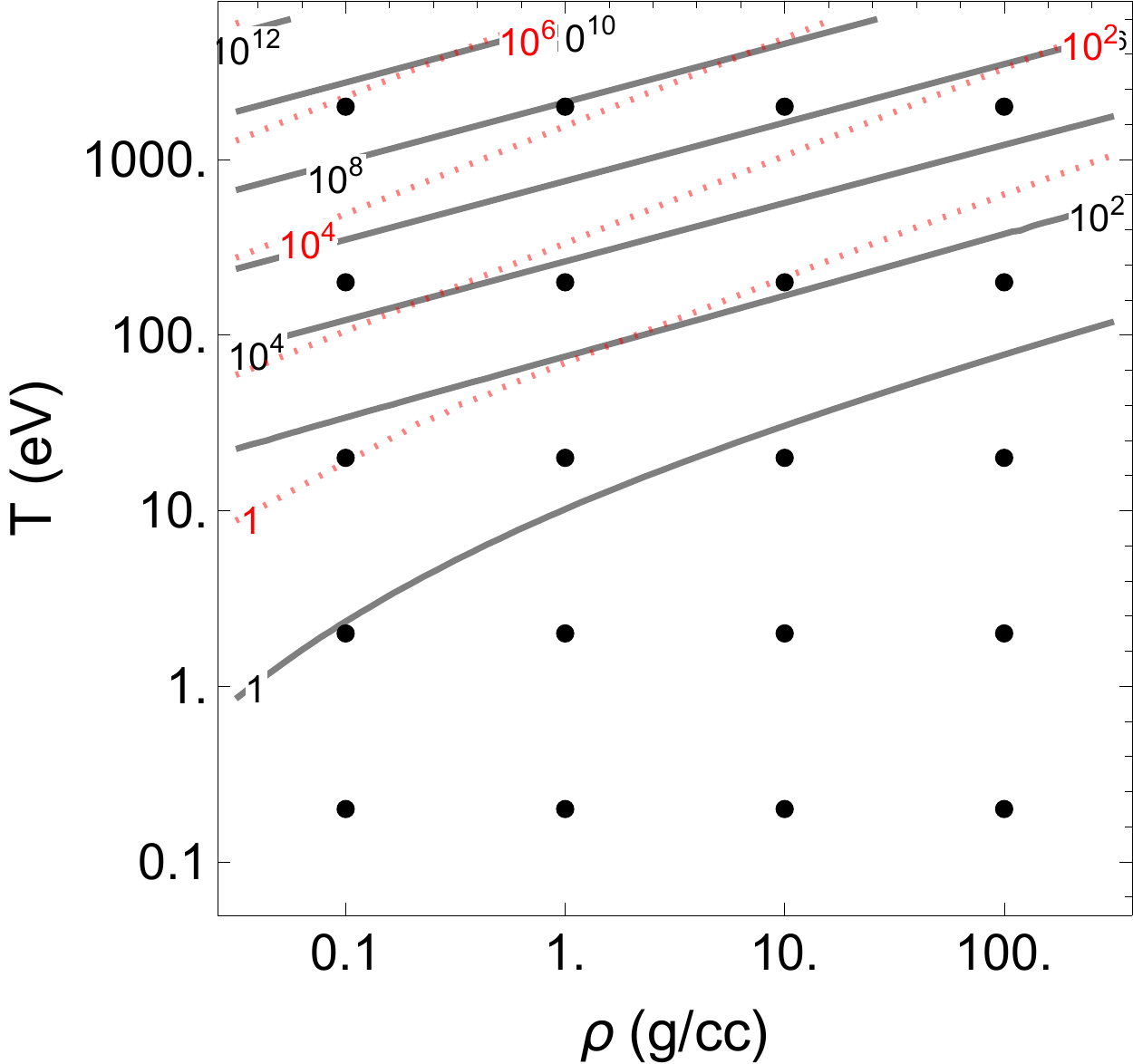}
\caption{\label{HVr} The ratio of $b_{max}^3/b_{min}^3$ (see text for definitions)
for ion-ion (solid black) and electron-electron (dotted red) interactions for 
hydrogen. This ratio
is a measure of the computational complexity of first-principles methods.
The workshop cases are at the positions of the black dots. The ratio becomes large and 
it becomes more difficult to apply first principles methods in the same regime where simple
physics approximations become valid.}
\end{figure}

\begin{figure}
\centering
\includegraphics[width=.95\columnwidth]{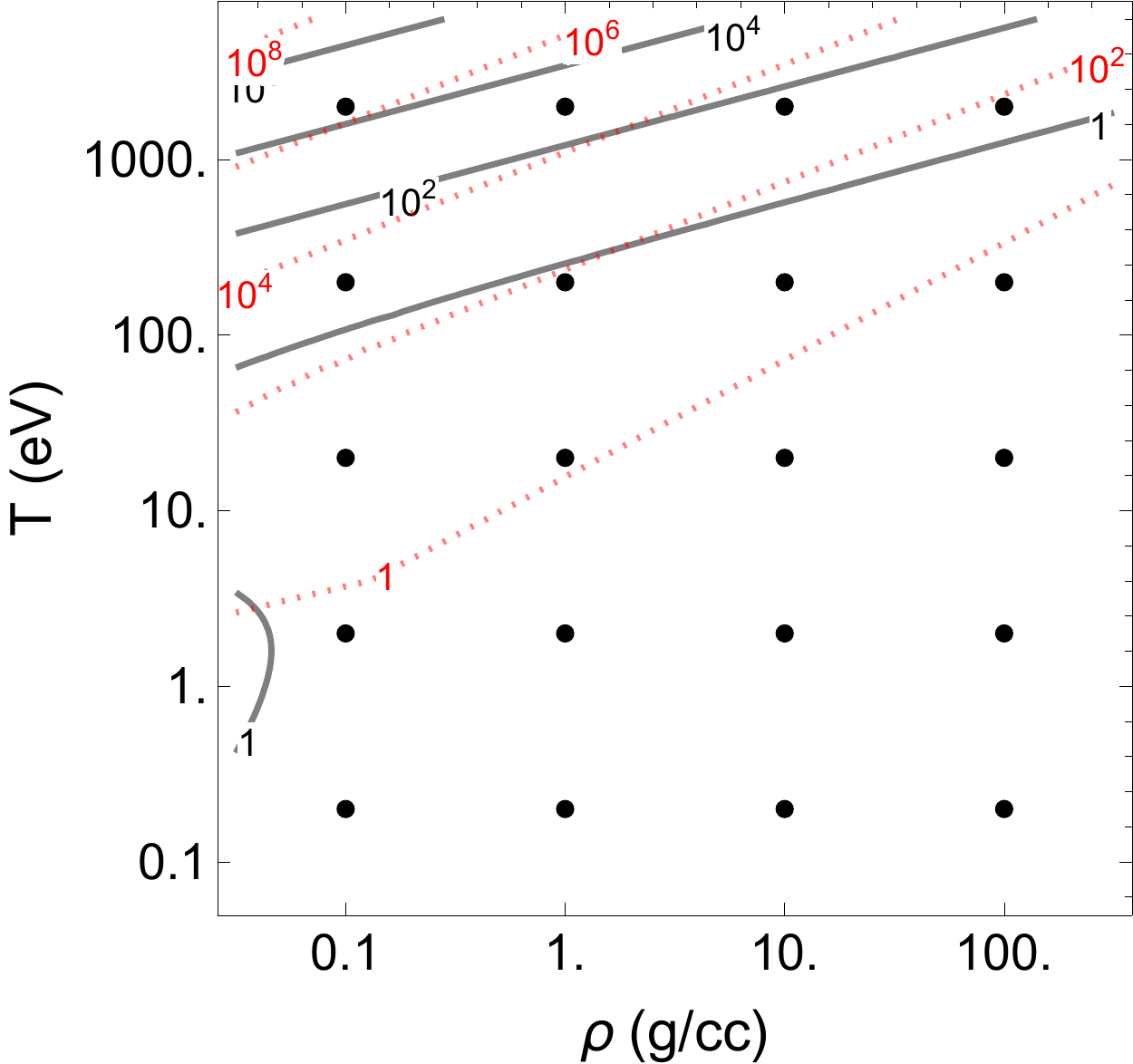}
\caption{\label{CVr} Same as Fig. \ref{HVr}, but for carbon.}
\end{figure}

Aside from the modeling challenges of nonideal plasma effects, there are also computational ones. A complete 
calculation should be converged with respect to system size while also resolving all important length scales. 
For example, a quantum molecular dynamics simulation should be large enough that ions are screened 
from their periodic images while also including enough basis functions to resolve the thermal de
Broglie wavelength as well as the atomic structure around each ion. We take as a rough measure of this 
complexity the ratio between the volume of a sphere of radius equal to the screening length to one 
with radius which is the maximum of the classical distance of closest approach and the thermal de Broglie wavelength:
\begin{equation}
 C_j= \frac{b_{j,max}^3}{b_{j,min}^3}.
 \end{equation}
The maximum impact parameter is given by a representative screening length. Since electrons are usually much faster than the ions, they tend to only be screened by
themselves, while ions are screened by both other ions and electrons. This distinction does
not apply to static properties involving a long-time average ($\omega=0$). However, for use
in dynamic contexts shorter than ion-motion time scales,  we approximate the maximum impact
 parameter by

 \begin{equation}
 b_{j,max}=\left\{\begin{array}{ll}
 \lambda_{TF} & \quad \textrm{if} j = \textrm{electron}\\
 \left(\lambda_{TF}^{-2}+\lambda_{DH,j}^{-2}\right)^{-1/2} &\quad \textrm{if} j = \textrm{ion}
 \end{array}\right.
\end{equation}
 where 
 \begin{equation}
 \lambda_{TF}^{-2}= 4\pi e^2 \frac{\partial n_e}{\partial \mu}
 \end{equation}
 is the Thomas-Fermi screening length
 and
 \begin{equation}
 \lambda_{DH,j} ^{-2}=4\pi q_j^2\frac{ n_j }{T}.
 \end{equation} 
 is the \old{Debye-Huckel} \new{Debye-H\"uckel} screening length for species $j$.
 The minimum impact parameter is given by the maximum of the thermal de Broglie wavelength and 
 the classical distance of closest approach:
 \begin{equation}
 b_{j,min}=\max\left[\sqrt{\frac{2\pi \hbar^2}{m_j T_{jj,eff}}}, \frac{q_j^2}{T_{jj,eff}}\right].
 \end{equation}
These types of length scales are usually the starting point for simple Coulomb logarithms (e.g. Landau-Spitzer 
\cite{Landau1936,Landau1937,Spitzer1962} or Gericke-Murillo-Schlanges \cite{GerickeEtAl2002}).
We plot the numerical complexity, $C_j$ for electrons and ions in pure hydrogen and carbon plasmas in 
Figs. \ref{HVr} and \ref{CVr}, respectively. This number gets very large in the high-temperature low-density limit
for both electrons and ions. Of course,
many things simplify at high temperatures; so simpler models become more valid and many details are
washed out. This measure should only
apply if one tries to do a brute force calculation, resolving all states and length scales.

Since the $\Gamma$'s and $C$'s are large in different limits, it is very difficult to have a model which can 
span the entire range of parameter space and hence, we almost never have a ``best'' model to compare against
for every condition, which would require heroic efforts of theory and computation. Proper interpretation
of the results of this code comparison requires keeping these complexity measures in mind.

\section{Theoretical Origins of Transport Coefficients}
Transport processes in hydrodynamic equations are typically described using linearized flux models, in which the leading
order coefficient associated with each process is known as a ``transport coefficient". While the underlying symmetries
of a given system will determine the general form of hydrodynamic equations, the actual transport coefficients (along 
with any equation of state information) must be determined by the micro-physics at the particle scale. For this reason,
there are several approaches to calculating transport coefficients, each of which requires connecting micro-physics 
processes to a particular hydrodynamic model.  

\begin{figure}
\centering
\includegraphics[width=.95\columnwidth]{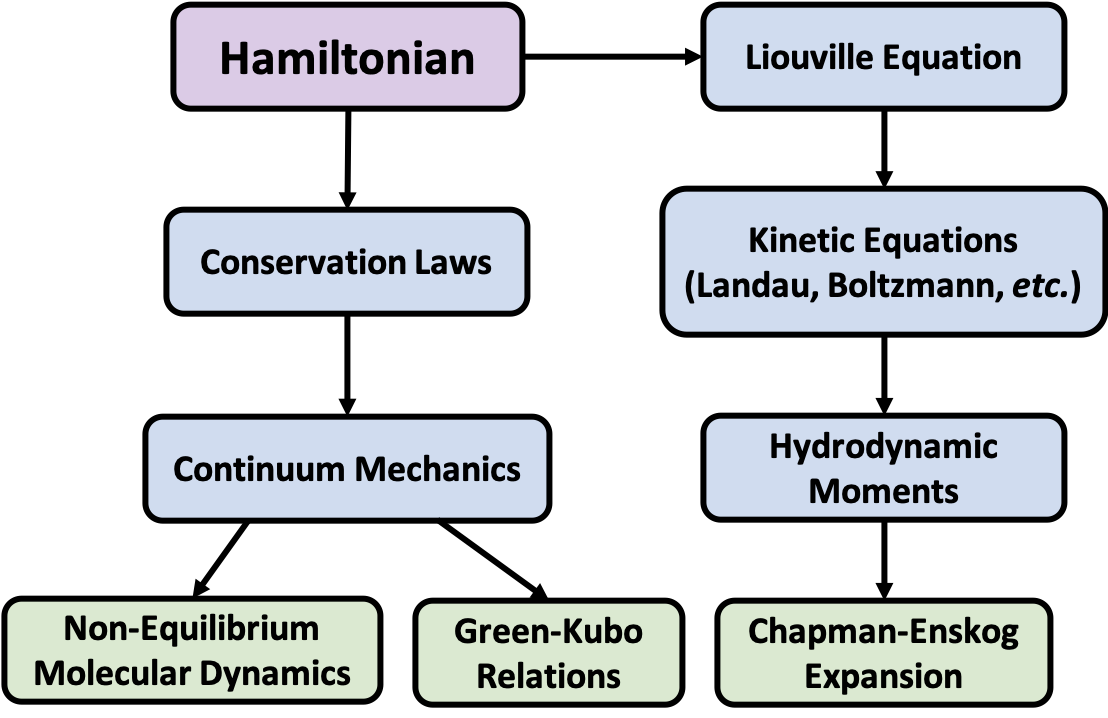}
\caption{\label{CPTCdiagram} Diagram of calculation methods for transport coefficients. In the left branch, transport
coefficients are calculated directly from correlations or currents of many-body simulations, while in the right branch, a 
kinetic equation is used to generate the continuum model and thus the corresponding transport coefficients.}
\end{figure}

We illustrate the most popular strategies of calculating transport coefficients in Figure (\ref{CPTCdiagram}), where we have 
categorized the methods into two main branches. The first branch, as discussed in Section (\ref{TCtheoryNEMD}), begins with a hydrodynamic 
model determined by symmetries and conservation laws, and micro-physics calculations are used to determine the transport 
coefficients.  Meanwhile, the second branch, as discussed in Section (\ref{TCtheoryCE}), uses a kinetic equation to derive a 
hydrodynamic model, and connects transport coefficients to micro-physics quantities in the process.

\new{We emphasize that this discussion is not intended to be a comprehensive review. 
Rather, this article aims to summarize the approaches taken by self-selected contributors to the workshop. 
For the interested reader, we here provide some important general references on the following relevant topics: 
fluid dynamics \cite{ChapmanEtAl1990,landau2013fluid,PfefferieEtAl2017},
kinetic theory \cite{Liboff2006,Bonitz2015}, dense plasma theory \cite{Ichimaru1982,IchimaruEtAl1987,Ichimaru2019}, 
density functional theory \cite{PribramJonesEtAl2015,PribramJonesEtAl2016}, quantum Monte Carlo \cite{DornheimEtAl2018}, 
response functions \cite{IchimaruEtAl1985}, molecular dynamics \cite{GrazianiEtAl2012},
average atom models \cite{MurilloEtAl2013}, atoms in plasma environments \cite{MurilloWeisheit1998,WeisheitMurillo2006},
wave packet molecular dynamics \cite{GrabowskiEtAl2013b,Grabowski2014},
ion transport \cite{StantonMurillo2016}, diffusion \cite{Daligault2012,HaxhimaliRudd1,HaxhimaliRudd2,OhtaHamaguchi}, viscosity \cite{TanakaIchimaru1986,Murillo2008,FortovMintsev2013}, 
electrical and thermal conductivity \cite{IchimaruTanaka1985,Ropke1988,KitamuraIchimaru1995,Apfelbaum2011,KnyazevLevashov2014,ShafferStarrett2020}, 
and stopping power \cite{YanEtAl1985,PeterMeyerterVehn1991a,PeterMeyerterVehn1991b,WangEtAl1998,GerickeSchlanges1999,GerickeEtAl1996,Zwicknagel1999,Gericke2002,GerickeSchlanges2003,GrabowskiEtAl2013,SRIM}.
}

\subsection{Origin of Transport Coefficients from Conservation Laws}\label{TCtheoryNEMD}

The most fundamental approach to calculating transport coefficients is to compare directly with the equations of hydrodynamics
that have been determined by the symmetries and conservation laws of the system of interest. In its most generic form,
the conservation of particle number, linear momentum and energy for a continuum field with no external sources can be expressed respectively 
by the equations
\begin{align}
    &\frac{\partial n}{\partial t} + \nabla\cdot\left(n{\bf v}\right) = 0,\\
    &mn\left(\frac{\partial{\bf u}}{\partial t} + {\bf u}\cdot\nabla{\bf v}\right) - \nabla\cdot\boldsymbol{\sigma} = 0,\\
    &n\left(\frac{\partial e}{\partial t} + {\bf u}\cdot\nabla e\right) - 
    \boldsymbol{\sigma}:(\nabla{\bf u}) + \nabla\cdot{\bf q} = 0.
\end{align}
Here, $m$ is the particle mass, $n({\bf r},t)$ is the number density, ${\bf u}({\bf r},t)$ is the velocity field, $e({\bf r},t)$ is the energy 
density, $\boldsymbol{\sigma}({\bf r},t)$ is the Cauchy stress tensor, and ${\bf q}({\bf r},t)$ is the heat flux.  Furthermore,
angular momentum is conserved by enforcing the symmetry $\boldsymbol{\sigma} = \boldsymbol{\sigma}^T$.  Relevant approximations
and closure relations are required to reduce this system of equations to the more familiar hydrodynamic models (Euler, Navier-Stokes,
{\it etc.}). 

For example, if we consider a binary system, we can introduce the number densities $n_{1,2}({\bf r},t)$ and velocity fields ${\bf u}_{1,2}({\bf r},t)$
for the individual species, each of which will satisfy a continuity equation associated with conservation of particle number
\begin{align}
    \frac{\partial n_i}{\partial t} + \nabla \cdot (n_i{\bf u}_i) = 0,\quad i = \{1,2\}.
\end{align}
In hydrodynamic models, diffusion is usually measured relative to the bulk motion of the fluid. For a binary fluid we can define the bulk center of 
mass velocity ${\bf u} = x_1 {\bf u}_1 + x_2 {\bf u}_2$, where $x_i = m_in_i/\rho$ and $\rho = (m_1n_1 + m_2n_2)$. The continuity equations are transformed 
into a frame moving at bulk velocity ${\bf u}$ to obtain
\begin{align}\label{continuity}
    \rho\frac{\partial x_i}{\partial t} + \rho {\bf u}\cdot \nabla x_i = -\nabla \cdot {\bf j}_i,
\end{align}
where ${\bf j}_i$ is the relative (to ${\bf u}$) mass flux density of the $i^{\text{th}}$ species.
For small gradients in the concentration field, the temperature field $T({\bf r},t)$,\footnote{Here and throughout this review, we will
assume there is only a single temperature for the sake of simplicity.} and the total pressure pressure field $P({\bf r},t)$, 
the Taylor expansion of the flux density can be truncated to linear order as
\begin{align}
\label{flux}
    {\bf j}_i \approx -\rho D\left(\nabla x_i + \frac{k_T}{T}\nabla T + \frac{k_P}{P}\nabla P\right),
\end{align}
where we have defined $D$ as the diffusion coefficient, $k_TD$ as the thermal diffusion coefficient, and
$k_P D$ as the barodiffusion coefficient \cite{landau2013fluid}.  Of course, gradients in these fields will appear in the remaining hydrodynamic
equations as well, in which each term will have its own coefficient that must be calculated from the micro-physics.
We have illustrated this approach in the left branch of the diagram in Figure (\ref{CPTCdiagram}).  

One such method along this branch is non-equilibrium molecular dynamics (NEMD), where molecular dynamics simulations are used to measure the response of a system to gradients in the appropriate state variables (density, temperature, {\it etc.}). This is possible because of the simple linear relationship in (\ref{flux}). NEMD calculations can be particularly challenging due to not only the computational complexity of the simulations, but isolating a gradient in a particular observable without inducing gradients in other quantities is often impossible as well; that is, one wishes to remain deeply in the linear regime to avoid generating the other gradients, and this creates signal-to-noise issues. 

\subsection{Green-Kubo Relations}\label{TCtheoryGK}

For {\it linear} transport coefficients one can also employ equilibrium MD. In this approach, one derives transport coefficients from equilibrium time correlation functions to generate the celebrated Green-Kubo (GK) relations for the transport coefficients. In this subsection we will sketch how the GK relationships arise. 
We illustrate this approach using inter-diffusion again as an example, which is the key process for atomic-scale mixing. 

If we assume gradients in the temperature and pressure fields to be negligible, the flux density (\ref{flux}) of the $i^{\text{th}}$ species reduces to the well-known form of
Fick's Law defined in terms of the relative density, ${\bf j}_i = -\rho D\nabla x_i$. Upon linearization, the continuity equation in Equation (\ref{continuity}) can thus be written as
\begin{align}
\frac{\partial x_i}{\partial t} + {\bf u}\cdot \nabla x_i  = D\nabla^2 x_i.
\end{align}
Note that the Fick's law for interdiffusion is written in terms of the total fluid mass density $\rho$ and the fractional density $x_i$ of species $i$. The resulting equation is then solved as an initial value problem in Laplace-Fourier space, which allows the equation to be written in terms of a time correlation function. So far, there is no connection to the microscopic dynamics, as all manipulations are consistent with macroscopic/hydrodynamic definitions. Next, the ensemble average of the resulting time-correlation function is written in terms of its {\it microscopic} definition, connecting the transport coefficient $D$ to the phase-space trajectory of the microscopic many-body system, yielding the GK formula for inter-diffusion \cite{boercker1987interdiffusion}
\begin{align}
    D = \frac{x_1x_2}{S_{cc}(0)}\int_0^\infty dt\,V_D(t).
\end{align}
Here, the autocorrelation function is defined as 
\begin{align}
    V_D(t) &\equiv \frac{1}{3Nx_1x_2}\langle{\bf v}_d(t)\cdot{\bf v}_d(0)\rangle,\\
    v_d(t) &\equiv x_2\sum_{i=1}^{N_1}{\bf v}_i(t) - x_1\sum_{j=1}^{N_2}{\bf v}_j(t),
\end{align}
where the first sum is over species 1, the second sum is over species 2, and $(\cdot)$ denotes a dot product between the two vectors. 
Furthermore, we have introduced the concentration structure factor, which is defined in terms of
the partial static structure factors as 
\begin{align}
    S_{cc}(k) = x_1x_2\left[x_2S_{11}(k)+x_1S_{22}(k)-2\sqrt{x_1x_2}S_{12}(k)\right].
\end{align}
Importantly, note how the resulting GK relation is intimately connected with specific choices and definitions at the hydrodynamic level. 
That is, each correlation function corresponds to a specific type of current with a precise definition, which should match
what is meant in the hydrodynamic equations.
In a similar fashion, one can employ the momentum equation of the Navier-Stokes equation to find a correlation function associated with viscosity that can be connected to microscopic dynamics; this strategy can be used for all of the other coefficients, and is readily adapted to quantum systems \cite{zwanzig1965time, mcquarrie2000statistical}.

While the GK relations present an elegant connection between transport processes and the underlying statistical 
mechanics of the system, they still leave the correlation functions themselves to be determined.  However, for the past few decades, there are a variety of computational methods, such as equilibrium molecular dynamics simulations, that can inform these correlation functions through detailed calculation of the trajectories. 

\subsection{Kubo-Greenwood and Ziman approaches}

The usual GK formulation can be connected with the  Boltzmann equation (under certain
 assumptions, see Ref. \cite{KadanoffBaym1962}) and this expresses  the static conductivity
 in terms of scattering cross sections and one-body distributions
instead of the current-current correlation function. In practice the
evaluation of the current-current correlation function can also be simplified by
 using the Fermi-Golden rule and the assumption of a momentum relaxation time ($\tau$)
 to calculate a conductivity.
The frequency-dependent  Kubo-Greenwood approaches uses Kohn-Sham eigenstates
 as the initial and final states of the scattering process to obtain a dynamic conductivity
 $\sigma(\omega)$, but the extraction of a static conductivity involves obtaining
 a $\tau$ by fitting to a Drude model, with the following formula for the static conductivity:
\begin{equation}
\label{CondEq}
\sigma=\frac{n e^2\tau}{m_e}
\end{equation}
in a standard notation. That is, in spite of the complexity of the theories, they all finally
 depend on the above equation with all its assumptions to extract the static conductivity.

 Equation \ref{CondEq} is also used in the Ziman formula. However, the Ziman formula does
 NOT use the static {\it current-current} correlation
 function, but relies on evaluating the static {\it force-force} correlation function. It gives
 the inverse of the conductivity (i.e., resistivity). In most applications, the result is
 equivalent to the use  of a Fermi golden rule with the initial and final states being plane
 waves, while the  scattering potential is a linearly screened weak pseudopotential.
 The evaluation of the inverse of the conductivity $R=1/\sigma$  rather than $\sigma$ is
 claimed to sum a larger class of scattering graphs, and the Ziman formula
 had been the preferred method for computations in liquid  metals (e.g., see Ref. \cite{Rossiter1987}).
 In fact Pozzo et al's computationally very heavy {\it tour de force} evaluation
 of the static conductivity of  liquid sodium using the Kubo-Greenwood formula, can be
 compared with easy calculations from the Ziman formula \cite{PozzoEtAl2011}.
 
 The Ziman formula is the favorite route in models based on the neutral-pseudo-atom (NPA)
 model \new{\cite{Ziman1964}}, or various types of average atom (AA) models. The NPA provides a free-electron pile up 
 $\Delta n(r)$  around the nucleus based on a Kohn-Sham calculation, and this is used to
 construct an electron-ion  pseudopotential $U_{ei}$ and an ion-ion pair potential $V_{ii}$.
 The latter is used in the hypernetted chain equation (with bridge corrections where needed) to generate
 an ion-ion structure factor $S_{ii}(k)$. The structure factor and the pseudopotential
 $U_{ei}(k)$ are used in the  Ziman formula. This ensures that the Kohn-Sham calculation, the
 pseudopotential, pair-potential and the structure factor are, in principle, self-consistent
 with each  other.

\subsection{Kinetic model based approaches}\label{TCtheoryCE}
As an alternative to calculating transport coefficients directly from equilibrium or non-equilibrium many-body simulations,
 kinetic equations ({\it e.g.}, Boltzmann,
Landau/Fokker-Planck, BGK) can be used to generate a hydrodynamic model through a hierarchy of hydrodynamic moments. While the hydrodynamic model 
will be limited by the often restrictive approximations of the governing kinetic equation, this approach has the advantage of analytic
simplicity over the methods listed in Section (\ref{TCtheoryNEMD}), which usually require a numerical many-body calculation.
We have illustrated this alternative approach in the right branch of the diagram in Figure (\ref{CPTCdiagram}).

We now demonstrate this approach using a very simple model. Suppose our kinetic equation has the BGK form \cite{Liboff2006}
\begin{align*}
    \frac{\partial f}{\partial t} + {\bf v}\cdot\nabla f = \frac{f_0 - f}{\tau},  
\end{align*}
where $f = f({\bf r},{\bf v},t)$ is the one-body distribution function, and $f_0 = f_0(n,{\bf u},T)$ is an equilibrium distribution function 
({\it e.g.}, a drifting Maxwellian) in terms of the density, mean velocity and temperature that has the same lowest-order 
moments as $f({\bf r},{\bf v},t)$, and $\tau$ is a collision time. The lowest-order moment yields the continuity equation for the density $n({\bf r},t)$, whereas 
the first-order moment yields the momentum equation for the fluid velocity ${\bf u}({\bf r},t)$
\begin{align*}
    &\frac{\partial n{\bf u}}{\partial t} + \nabla \cdot \langle {\bf v} {\bf v} \rangle = 0,\\
    &\frac{\partial n{\bf u}}{\partial t} + \nabla \cdot  {\bf u} {\bf u} + \nabla \cdot \langle {\bf c} {\bf c} \rangle = 0.
\end{align*}
In the second line the mean velocity has been separated to isolate the central moment in terms of the relative velocity ${\bf c} = {\bf v} - {\bf u}$.  
This equation is not closed as a result of the term $\langle {\bf c} {\bf c} \rangle = \int d^3v \: f {\bf c} {\bf c}$. If the collision time is very 
small, which corresponds to a very short mean-free path, the kinetic equation yields the approximate solution $f\approx f_0$; using $f_0$ in $\langle {\bf c} {\bf c} \rangle$ yields the usual pressure term in terms of the 
temperature, which leads to the Euler form of hydrodynamics. For weaker collisions, an improved solution of the kinetic equation is needed, of which the lowest-order solution is
\begin{align*}
    f \approx f_0 - \tau\left({\bf v}\cdot \nabla f_0 \right).    
\end{align*}
This is a simple form of the Chapman-Enskog (CE) expansion \cite{ChapmanEtAl1990}. 

The second term, when used to evaluate 
$\langle {\bf c} {\bf c} \rangle $, yields transport terms that are proportional to the collision time $\tau$. Note that the transport terms will contain 
terms proportional to $\nabla n$, $\nabla {\bf u}$ and $\nabla T$ because of the dependencies in $f_0$. Moreover, note that we can identify the various 
transport processes by inspection and the transport coefficient naturally arises as the coefficient of those terms. In this case, 
all of the transport coefficients would self-consistently be connected through $\tau$; other  kinetic equations, such as 
Landau/Fokker-Planck or Boltzmann, would have similar properties.

\subsection{Open Questions}\label{Openness}

While the results presented below pertain to the linear transport regime, we wish to mention some extensions worth future study. As we have seen,  linearization is used at nearly every step, either in writing the fluxes, or, equivalently, in keeping the lowest order terms in the CE expansion and in obtaining the GK relations. Non-linear contributions yield both higher-order terms in the gradients, but also generate cross terms. The cross terms couple the various transport processes to create new forms of transport; for example, the Biermann battery \cite{matteucci2018biermann} results from a $\nabla n\times \nabla T$ cross term. Moreover, it is possible in some cases for the non-linear terms to create a situation where the steady state flux relationships of the form ${\bf j} \sim -{\cal C} \nabla U$ no longer hold. In addition, as we will see below, most of the transport coefficients have their largest values, and therefore are more important to the hydrodynamic evolution, at higher temperatures (weaker coupling). At high enough temperature, the mean-free path of the particles will exceed the gradient scale length (e.g., the scale of the density gradient in diffusion); this is the non-local transport regime 
\cite{bychenkov1995nonlocal, gribnikov1995nonlocal} for which transport coefficients cease to have their usual utility; non-locality in transport is particularly important in thermal conduction \cite{Callen1997}. 

We have also limited ourselves to the canonical ensemble of fixed volume with one temperature for all species. 
However, the advent of
fast lasers and the ability to create two-temperature plasmas that can be probed using
femto-second pulses has led to the possibility of studying dynamic conductivities dependent on
two temperatures, that is with ion and electron temperatures such that $T_i\ne T_e$. Furthermore, laser techniques
allow the study of isochoric conductivities, whereas most techniques used up to the 1980s
(e.g., for liquid metals) were for isobaric conductivities. This distinction has not been
well understood  by the community, and it is not unusual to see papers where isochoric
conductivities are compared with isobaric conductivities \cite{DharmawardanaEtAl2017}.
 

\section{Models}

Many choices go into each model. Their biggest difference is whether the electrons are treated dynamically or statically.
A fully dynamic treatment allows an explicit calculation of time-dependent electron-ion and electron-electron correlations, which 
can be used in GK formulas \cite{Green1954, Kubo1957} or the quantity of interest (e.g. the energy of the projectile in stopping power)
can be directly tracked. Example time-dependent methods which were contributed to our workshop are Time-Dependent Density Functional Theory (TDDFT)
\cite{MagyarEtAl2016,AndradeEtAl2018} and electron Force Field (eFF) \cite{SuGoddard2007,SuGoddard2009}. 
Alternatively, a great simplification of electron dynamics can be made by modeling their effect with the Langevin equation \cite{DaiEtAl2010}.

All the other models rely on the Born-Oppenheimer approximation, in which the electrons are assumed 
to instantaneously screen the much slower ions. The ion motions from molecular dynamics yield the viscosity and diffusion,
while electron dynamics must be inferred through some approximations to the electron-ion and electron-electron
collision operators of kinetic theory.
Approximations to the electron screening differ most in how they treat ion-ion correlations.
For example, the average atom assumes the ion-ion radial distribution function is a step function at the ion-sphere
radius. The correlations can later be inferred by mapping the average atom to the one component plasma 
with an effective ion charge, such as in PIJ \cite{Arnault2013}, or altering the ion charge and introducing a screening length as in KMD \cite{HaxhimaliEtAl2015,StantonMurillo2016}.
 Using the quantum hypernetted chain approximation \cite{Chihara1973,Chihara1978,DharmawardanaPerrot1982} one can obtain spherically symmetric potentials that can exhibit liquid-like
behavior as in EPT \cite{BaalrudDaligault2013}, but angular correlations are neglected unless one does a three-dimensional
quantum mechanical density functional theory (DFT) \cite{PribramJonesEtAl2015} calculation.
Such DFT calculations come in several different flavors, all of which are based upon semiclassical 
approximations about uniform electron density of the electron free energy. 
In order of increasing expected accuracy and increasing computational cost, 
these are Thomas-Fermi  \cite{Thomas1927,Fermi1927}, Sjostrom-Daligault (SDMD) \cite{SjostromDaligault2014}, and Kohn-Sham \cite{KohnSham1965}.
The derived potentials can be used in two different ways:
effective binary interactions can be used within Boltzmann collision operators as in 
EPT \cite{BaalrudDaligault2013} or the binary or many-body interactions can be used in molecular dynamics simulations \cite{StarrettSaumon2016}.  

Whenever a model includes a sufficient representation of the electrons' state, quantities dependent on electron-ion or electron-electron collisions 
(e.g. electrical and thermal conductivities and stopping power) can be calculated. We employed four types of approximations: Ziman-type expressions 
\cite{Ziman1961,EvansEtAl1973}, which are used in various single-center plasma models \cite{PerrotDharmawardana1987,Dharmawardana2006,SterneEtAl2007,Rozsnyai2008,PainDejonghe2010,FaussurierBlancard2015,BurrillEtAl2016},
the Kubo-Greenwood approximation \cite{Greenwood1958}, which approximates the electron wave function with the Kohn-Sham wave function and
can be employed in either single-center models \cite{Johnson2006,KuchievJohnson2008,Johnson2009,StarrettEtAl2012} 
or three-dimensional multi-atom calculations \old{\cite{DesjarlaisEtAl2002},\cite{HansonEtAl2011},}
\new{\cite{DesjarlaisEtAl2002,HansonEtAl2011,HolstEtAl2011,FrenchEtAl2012,BeckerEtAl2018},}
local density approximations \old{\cite{SarasolaEtAl2001}} \new{\cite{SarasolaEtAl2001,WangEtAl1998}}, which use homogeneous electron gas formulas as a starting point for the inhomogeneous problem (almost all
stopping power models used this approximation \new{when bound electrons are important}), and directly simulating dynamics \cite{MagyarEtAl2016,BaczewskiEtAl2016,Correa2018}.


\section{Results}

In this section, we present five different quantities (electrical and thermal conductivity, viscosity, diffusion, and stopping power)
for three types of materials (hydrogen, carbon, and an equimolar mixture of the two) across twenty different conditions
(four different densities, five different temperatures, and in the case of stopping power, seven different energies). 
There were up to 15 submissions for each of the included 520 cases (we had no submissions for stopping in the CH mixture).
Since one of our main goals is to facilitate sensitivity analyses, we focus on mean and spread measures. However, 
for the researcher interested in making detailed comparisons, we include some discussion below and 
have included all the submitted data as a supplementary data file, 
except from those who requested anonymity. 

It is useful to give a measure of the spread in the reported values for each quantity. A conservative measure would be the ratio of maximum to minimum values, but this 
tends to overemphasize outliers. We would also prefer a method that uses information from all of the data rather than the two extrema. A more natural quantity to assess is 
the standard deviation, $\sigma(\{d_i\})$, where $\{d_i\}$ represents our data set for a particular quantity at some condition. 
However, data which is spread over decades may give a value of $\sigma$ which is very skewed by one data point. 
Therefore, we define a spread measure
\begin{equation}
\label{sigmadef}
\tilde{\sigma}=\exp[\sigma(\{\ln d_i\})]-1.
\end{equation}
Note, this measure does not depend on the units or scale of the data. 
Small values of $\tilde{\sigma}$ (much less that one) mean that there is little variation while very large values indicate lack of consensus. These values correspond roughly to the fractional variation within the data.

\begin{figure*}
\centering
\includegraphics[width=.95\linewidth]{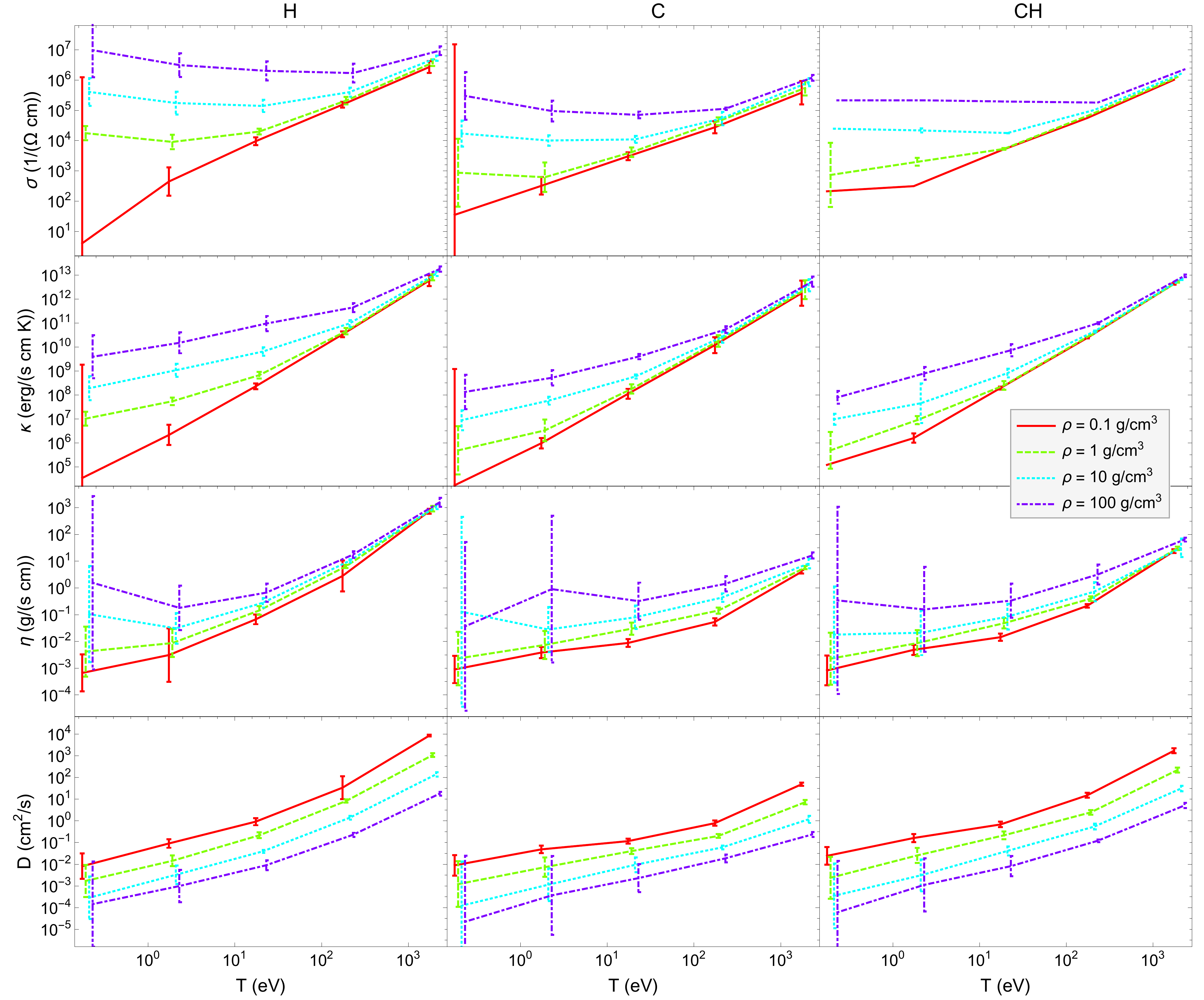}
\caption{\label{HAll} 
Electrical (first row) and thermal (second row) conductivity, ion viscosity (third row), and the diffusion coefficient (fourth row) for pure hydrogen (first column), pure carbon (second column), and an equimolar carbon-hydrogen mixture. For the pure cases
the diffusion coefficient is the self diffusion while for the mixture, it is the interdiffusion. 
Plotted are the mean (position) and standard deviations (error bars) in logarithmic space of all submissions as a function 
of temperature. The different density cases plotted are $\rho = 0.1\,$g/cm$^3$ (red, solid), $\rho = 1\,$g/cm$^3$ (green, dashed), 
$\rho = 10\,$g/cm$^3$ (cyan, dotted), and $\rho = 100\,$g/cm$^3$ (purple, dot-dashed). The spreads tend to be larger at low
temperatures or high densities \new{For clarity, the different densities curves are slightly offset from one another in temperature from the actual values of 0.2, 2, 20, 200, and 2000 eV.}
}
\end{figure*}
\begin{figure*}
\centering
\includegraphics[width=.95\linewidth]{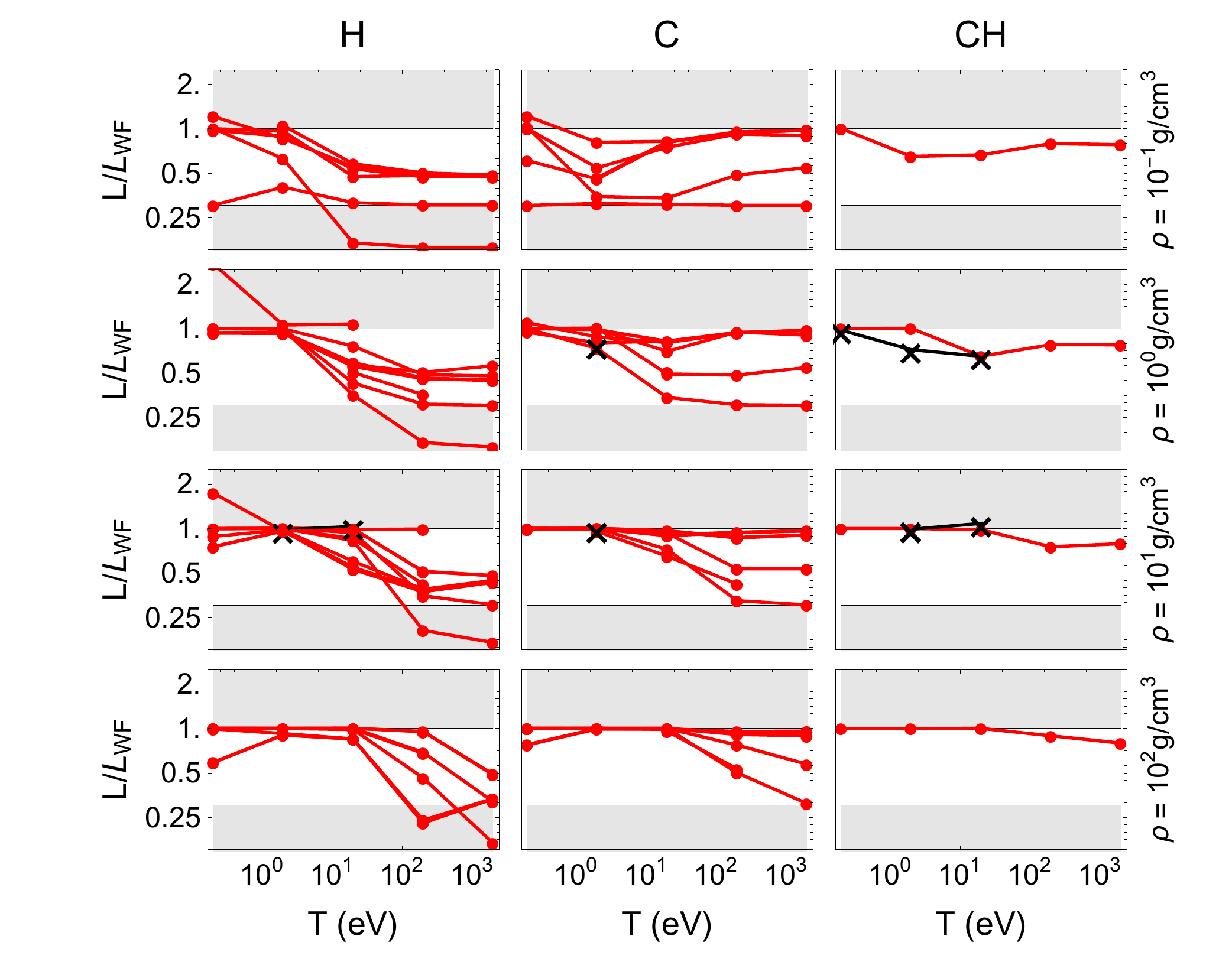}
\caption{\label{Lorenz} Lorenz number ($L=\kappa/(\sigma k_B T)$) divided by the Widermann-Franz limit ($L_{WF} = \pi^2 k_B^2/(3e^2)$) of all submitted calculations. The Wiedermann-Franz and Spitzer ($L_S = k_B^2/e^2$) limits are shown by the 
upper and lower shaded gray regions, respectively. All the data is shown as circles connected by solid red lines except for 
three dimensional density functional calculations (black x's), which were the most expensive.}
\end{figure*}

\begin{figure*}
\centering
\includegraphics[width=.95\linewidth]{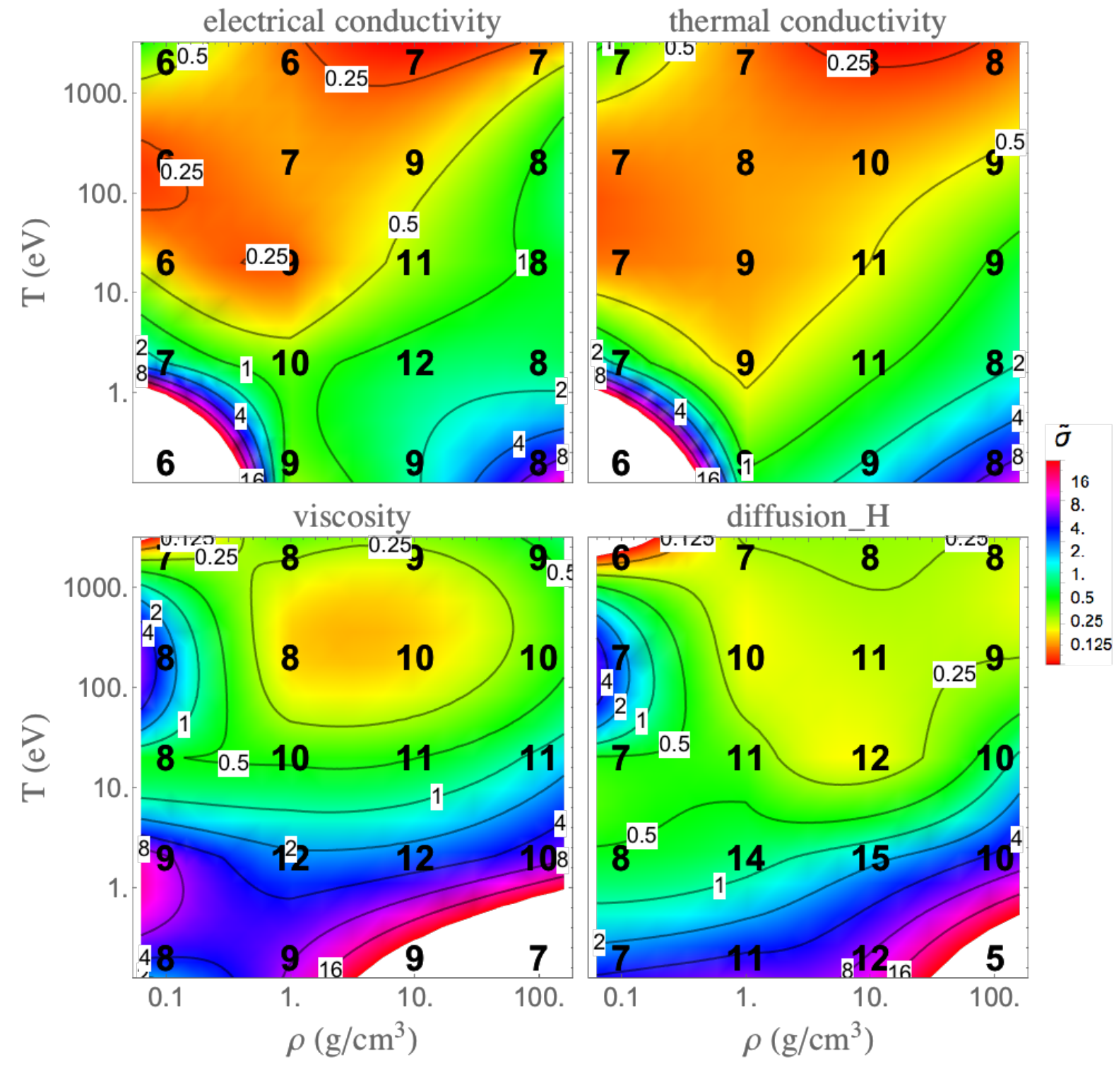}
\caption{\label{HSig} Spread in reported values of transport coefficients for hydrogen. This spread roughly 
corresponds to the fractional variation in the submitted values. See Eq. \ref{sigmadef} for the definition. 
The large black bold numbers
are located at the positions of the workshop cases and their values are the number of submissions at those conditions. Small values (red) indicate agreement among submissions while large values (magenta) indicate disagreement. 
The larger values correspond to when either 
the relevant Coulomb coupling parameter or computational expense is large (see Sec. \ref{SecDimensionless}).}
\end{figure*}

\begin{figure*}
\centering
\includegraphics[width=.95\linewidth]{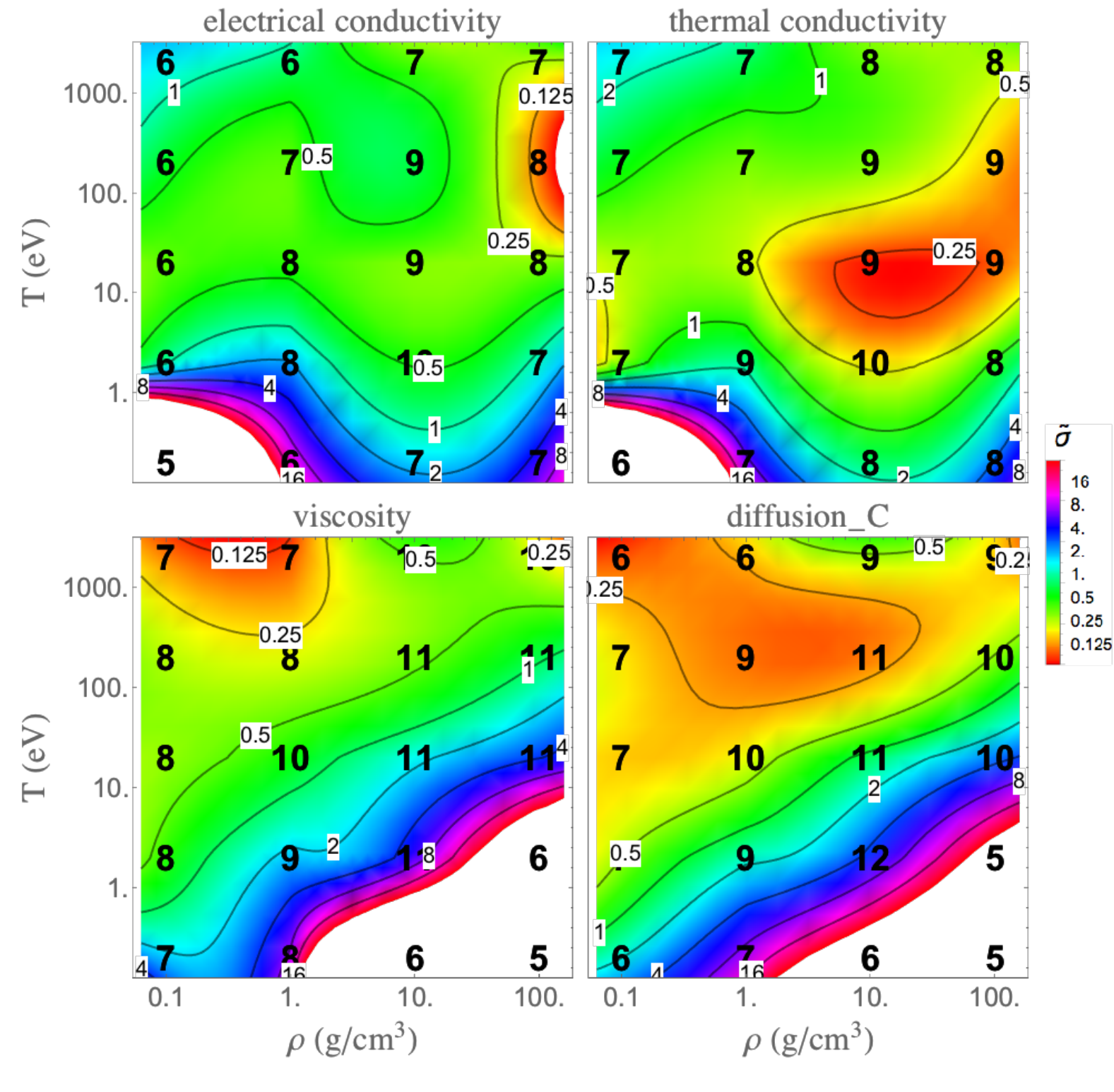}
\caption{\label{CSig} Same as Fig. \ref{HSig} but for carbon.}
\end{figure*}

\begin{figure*}
\centering
\includegraphics[width=.95\linewidth]{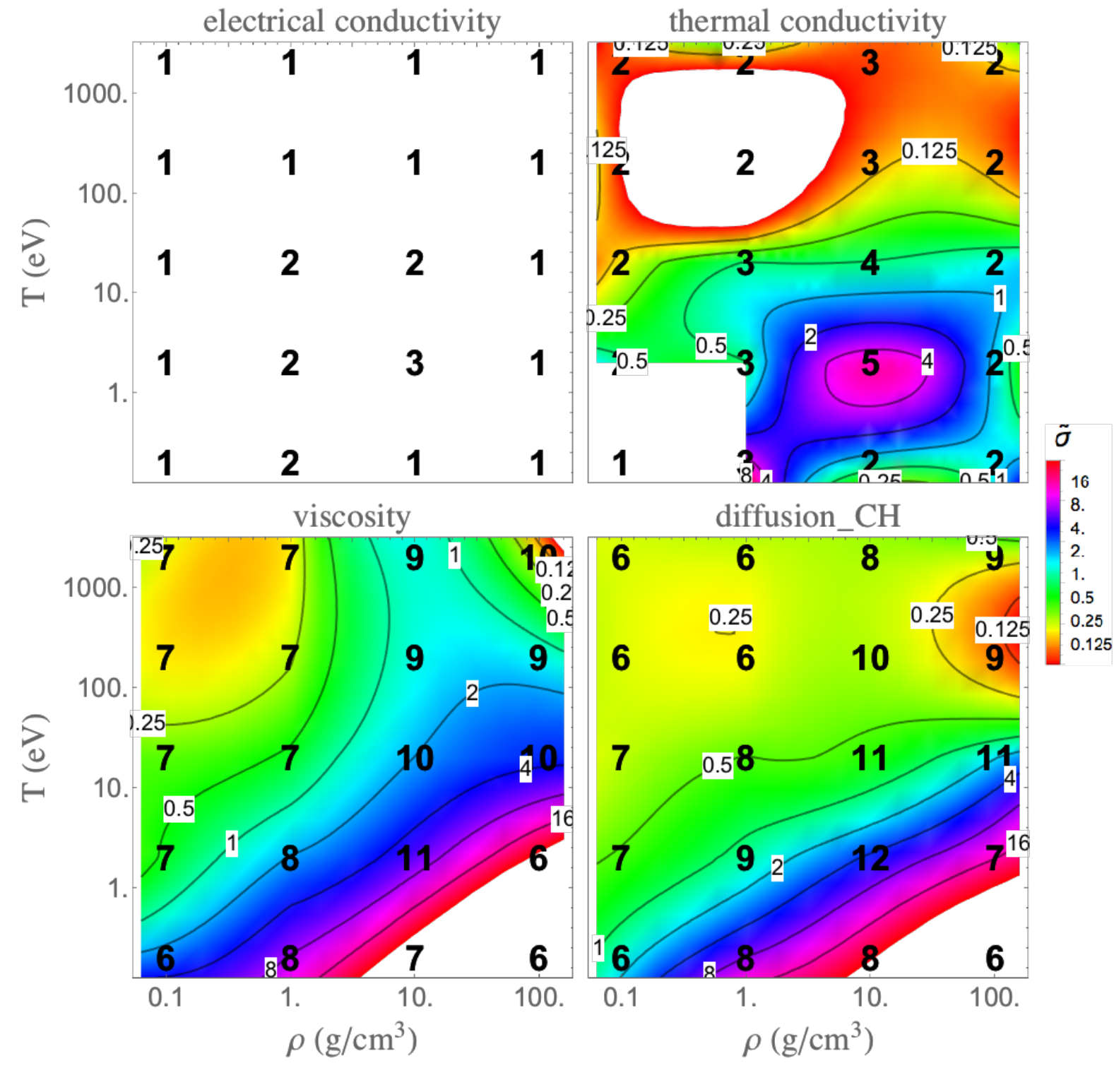}
\caption{\label{CHSig} Same as Fig. \ref{HSig} but for the equimolar carbon-hydrogen mixture. Note, there were not enough 
electrical conductivity submissions to report a valid spread measure. The reliability of the thermal conductivity spread measure 
here is also marginal due to having too few submissions.}
\end{figure*}


\subsection{Electronic Transport}

The means and standard deviations of all submissions  of electrical and thermal conductivity are plotted in Fig. \ref{HAll}. 
Individual submissions are listed in the appendix. 
For all materials, we find fair agreement among the four independent average-atom models for both electrical and thermal conductivities. These models also agree fairly well with the MD models at moderate densities. The outliers tend to be the parameterized analytic models, while values actually used in hydrodynamic simulations tend to fall within the range of more sophisticated models. For carbon, the widely used \old{Lee-more} \new{Lee-More} approximation agrees with QMD at 1 g/cc but is a factor of 10 too small at 10 g/cc. Large disagreements at low temperatures and densities are due primarily to differences in calculated ionization, while large disagreements at high densities are attributable to differing treatments of ionic structure and electron degeneracy. For hydrogen in the high-density regime relevant to ICF stagnation, factors of three to ten among models persist even at high temperatures.  These spreads can be 
seen in Figs. \ref{HSig} to \ref{CHSig}.

The Lorenz number is shown in Fig. \ref{Lorenz}. For both carbon and hydrogen, most codes recover Weideman-Franz scaling 
\cite{Kittel2018} at high densities and low temperatures. There is less agreement among calculated thermal to electrical conductivity ratios in the classical limit, where the treatment of electronic collisions in both electrical and thermal conductivities varies among the models.


\subsection{Ionic Transport}

Ion viscosity and ion diffusion coefficients are presented in the lower two rows of Fig. \ref{HAll}. In the pure element
cases, the self-diffusion coefficient is shown, while for the mixture, it is the interdiffusion coefficient. 
Because ions are much more massive than electrons, they carry most of the momentum in the plasma. 
Consequently, viscosity is associated with ions. 
The electric field associated with electrons may influence ion-ion interactions \new{\cite{Murillo2008}}, but the electrons themselves do not provide a significant direct contribution. 

A striking feature of the shear viscosity coefficient at warm dense matter conditions is that a minimum can be reached as temperature varies. 
This is associated with the underlying physical mechanisms responsible for viscosity, which are directly illustrated in the Green-Kubo relation \cite{HansenMcDonald2006}. 
At weak coupling (low density or high temperature), shear viscosity is determined by the kinetic energy of particles. 
In this regime, traditional Landau-Spitzer \cite{Landau1936,Landau1937,Spitzer1962} theory predicts that the shear viscosity increases with temperature as $\eta \propto T^{5/2}$. 
Essentially all of the submissions appear to capture this limit. 
At sufficiently low temperature, the kinetic energy of particles does not contribute significantly, and the shear viscosity is instead determined by the Coulomb potential energy of ions. 
Much understanding of this transition is based on the one-component plasma model, where it occurs at $\Gamma \simeq 17$ \cite{SaigoHamaguchi2002,DonkoHartmann2008,DaligaultEtAl2014}. 
In low temperature dense plasmas, shear viscosity decreases with increasing temperature, as it does in ordinary liquids. 
This is a challenging regime for theory because the many-body potential energy is difficult to model. 
It is also a challenging regime for first-principles simulations because ion transport occurs much more slowly than electron transport, which often requires unattainably long simulations to achieve reliable results.

As expected, the spread in data at high density low temperature conditions is especially severe, often spanning several orders of magnitude. 
At these conditions there is little agreement among any of the submissions, illustrating that this is one of the least understood transport processes considered in this workshop. 
Ion viscosity in this regime is relevant to modeling the fuel-shell interface in ICF (e.g. Ref. \cite{RosenbergEtAl2015}) 
This highlights one of the most important areas where improved theory and simulations are needed. 
It may contribute to a high degree of uncertainty for some aspects of hydrodynamic simulations when these conditions are encountered. However, it is important to recall that small values of transport coefficients imply correspondingly small contributions to the hydrodynamic equations: the stronger the collisions, the smaller the transport coefficients, and the smaller the terms in the hydrodynamics model, unless there are extremely large gradients. 

The submitted ion diffusion coefficients tend to be in closer agreement than ion viscosity coefficients, but with still substantial disagreement in the strongly coupled regime. 
Analytic models tend to be outliers.  
Again, the agreement is best at high temperature and low density conditions (weak coupling), where all submissions appear to asymptote to the $D \propto T^{5/2}$ regime of Landau-Spitzer theory. 
Similar trends are observed for the self-diffusion processes in the single component systems as for interdiffusion in the  CH mixture. 
With the exception of one analytic model, all submissions predict that $D$ monotonically decreases with decreasing temperature, consistent with expectations from the one-component plasma \cite{HansenEtAl1975,OhtaHamaguchi2000}. 
The better agreement for diffusion, compared to viscosity, may be that the correlation function for diffusion is entirely determined by the kinetic energy of particles. 
Thus, there is not a fundamental transition in the physical mechanism responsible for diffusion as there is with viscosity. 

Diffusion processes are important in ICF plasmas particularly with regard to deuterium and tritium fuel mixing, or demixing, near a hot spot, but also in the mixing of shell materials near the edge of the fuel. 
The data suggest that reliable models exist in the weakly coupled plasma regimes, such as may be expected in the former example of fusion fuel mixing in a hot plasma. 
However, there may be much less reliable models in the latter example concerning mixing of shell materials in the cooler outer regions of the plasma. 
Continued progress in such simulations will rely on further improvements to the diffusion models particularly in these more dense or cool regions. 


\subsection{Stopping Power}

Accurate values for the stopping powers are needed for ICF target design \cite{ZylstraHurricane2019} 
because alpha particle heating maintains the temperature in the presence of energy loss mechanisms. 
Warm dense matter experiments can also rely on charged particle beams for heating \cite{KimEtAl2015,BangEtAl2015,McKelveyEtAl2017} or as a probe \cite{ZylstraEtAl2015,FrenjeEtAl2015}. 
Unlike the electronic and ionic transport properties, which represent an integration over a distribution (usually thermal) 
of charged particle velocities, stopping powers must be energy-resolved to track the thermalization of fast fusion 
products through collisions with the background material.
Hence, they more distinctly probe different parts of the underlying collision operator needed for all transport 
quantities while also having an extra dimension of parameter space to explore. 

\new{Over the last few decades there have been many different stopping power experiments in plasmas. These
can be divided into experiments involving high-energy, moderate to heavy ions
in weakly coupled plasmas \cite{JacobyEtAl1995,HoffmannEtAl1990,RothEtAl2000,FrankEtAl2013,CayzacEtAl2017},
moderate-energy, fully-ionized, light ions in weakly coupled plasmas 
\cite{HicksEtAl2000,FrenjeEtAl2015,ChenEtAl2018,SayreEtAl2019,FrenjeEtAl2019}, high-energy, fully-ionized, light ions in weakly coupled plasmas \cite{HayesEtAl2015} 
moderate-energy, light ions in a moderately coupled plasma \cite{GrazianiEtAl2012}, and high-energy, light ions in a moderately coupled plasma \cite{ZylstraEtAl2015,ZylstraEtAl2016}, where high (moderate) energy refers to ions with velocities significantly greater than 
(approximately equal to) the thermal velocity of the target electrons. While all of these experiments are very useful for 
benchmarking theoretical predictions, it remains difficult to produce mono-energetic projectiles in a temperature and density
gradient-free dense plasma and precisely measure energy losses. Such measurements remain sparse in the high-dimensional
space of projectile energy, projectile charge, density, temperature, and target material. For example, we are not aware of any such precision
measurements of moderate-energy ions in warm dense matter.}

\begin{figure}
\centering
\includegraphics[width=.95\linewidth]{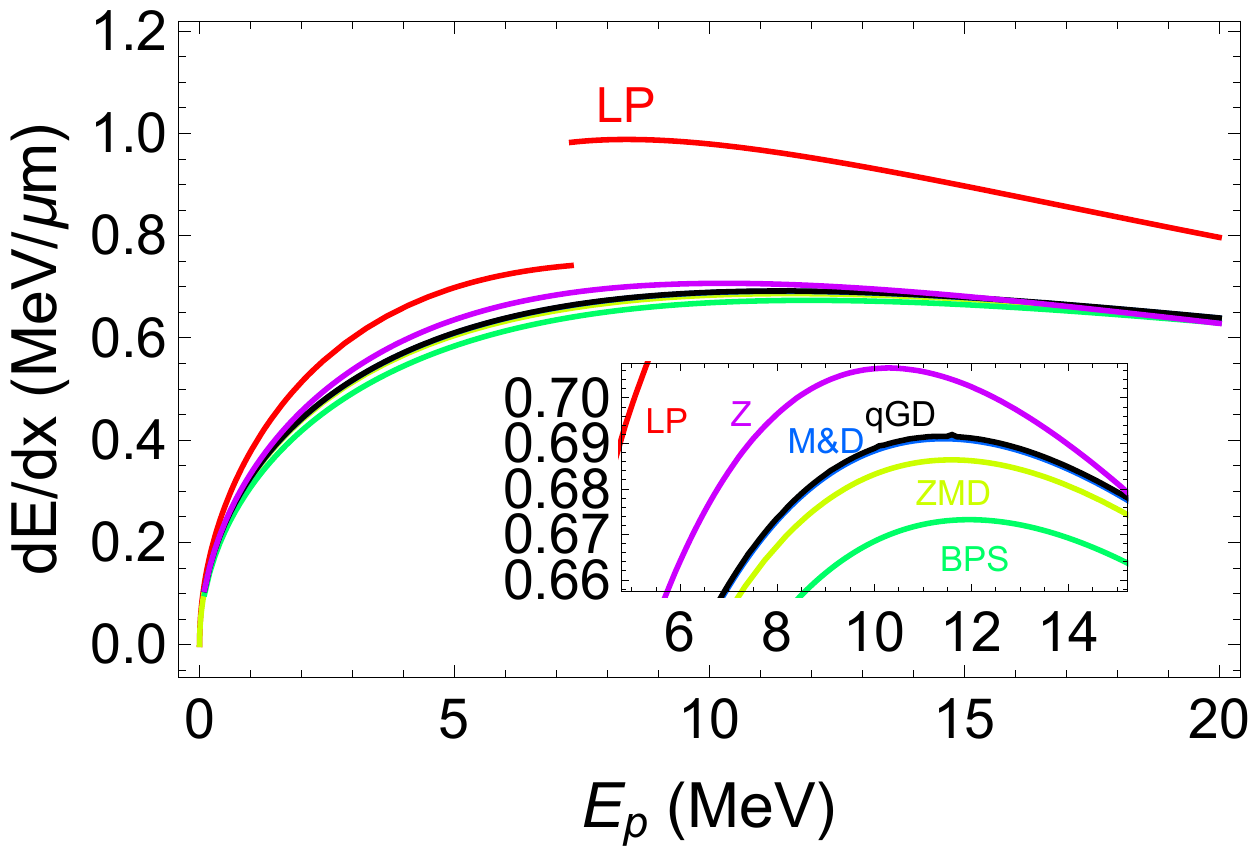}
\caption{\label{ESP} Stopping power of an alpha particle in a uniform electron gas as a function of projectile energy at $n_e = 
10^{25}\,$cm$^{-3}$ and $T = 1\,$keV for the Li-Petrasso (LP) \cite{LiPetrasso1993}, Zwicknagel (Z) \cite{Zwicknagel1999}, 
Brown-Preston-Singleton (BPS) \cite{BrownEtAl2005}, the quantum Gould-DeWitt (qGD) \cite{GouldDeWitt1967}, Maynard-Deutsch (M\&D) \cite{MaynardDeutsch1985}, and Zimmerman's fit
to Maynard-Deutsch (ZMD) \cite{Zimmerman1990} models . At these weakly coupled conditions, all models shown accurately produce the 
high energy limit (with LP doing so by adding a step function correction for higher energies). Two models, M\&D and qGD 
are indistinguishable, indicating that strong scattering effects are negligible at these conditions. }
\end{figure}

\begin{figure}
\centering
\includegraphics[width=.95\linewidth]{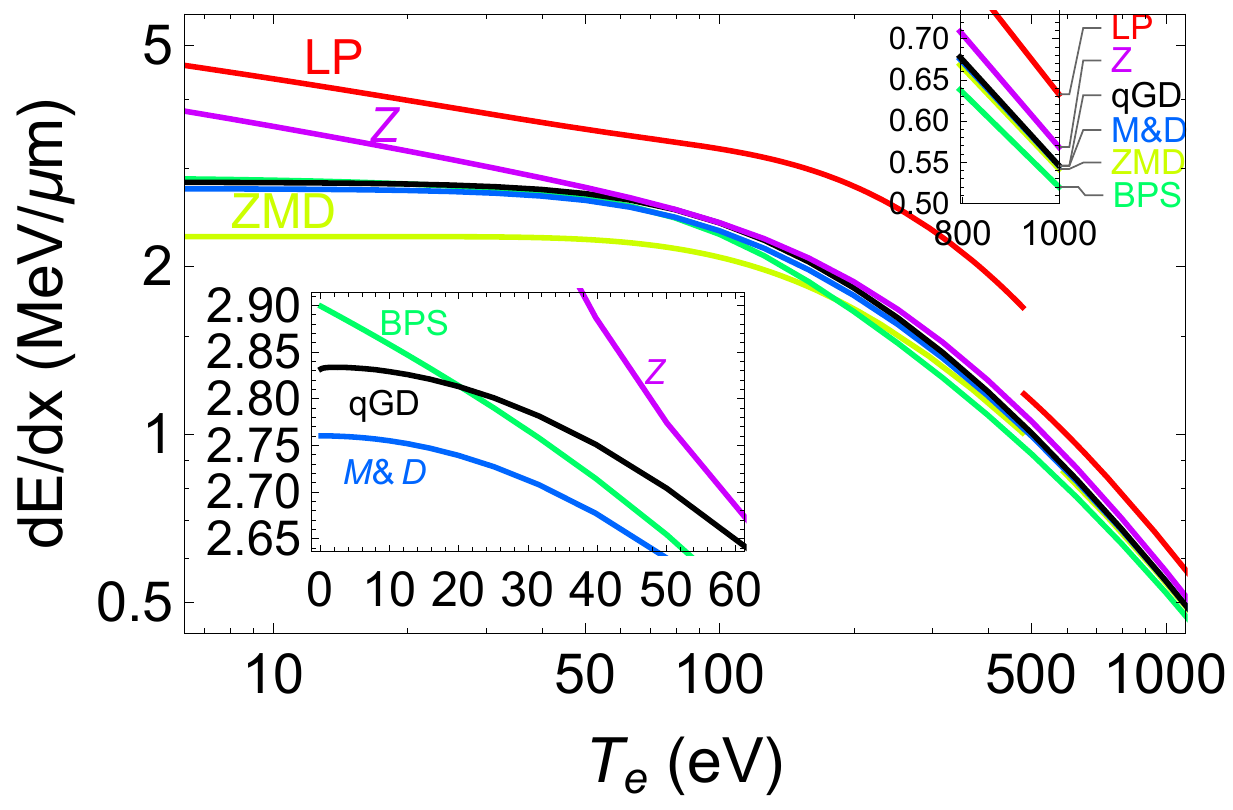}
\caption{\label{TSP} Same as Fig. \ref{ESP} but fixing the energy at $3.5\,$MeV and varying the temperature. We see that the Z
model as incorrect behavior at low temperature and Zimmermans fit (ZMD) is off by about 18\% with respect to M\&D in this limit.}
\end{figure}

\begin{figure}
\centering
\includegraphics[width=.95\linewidth]{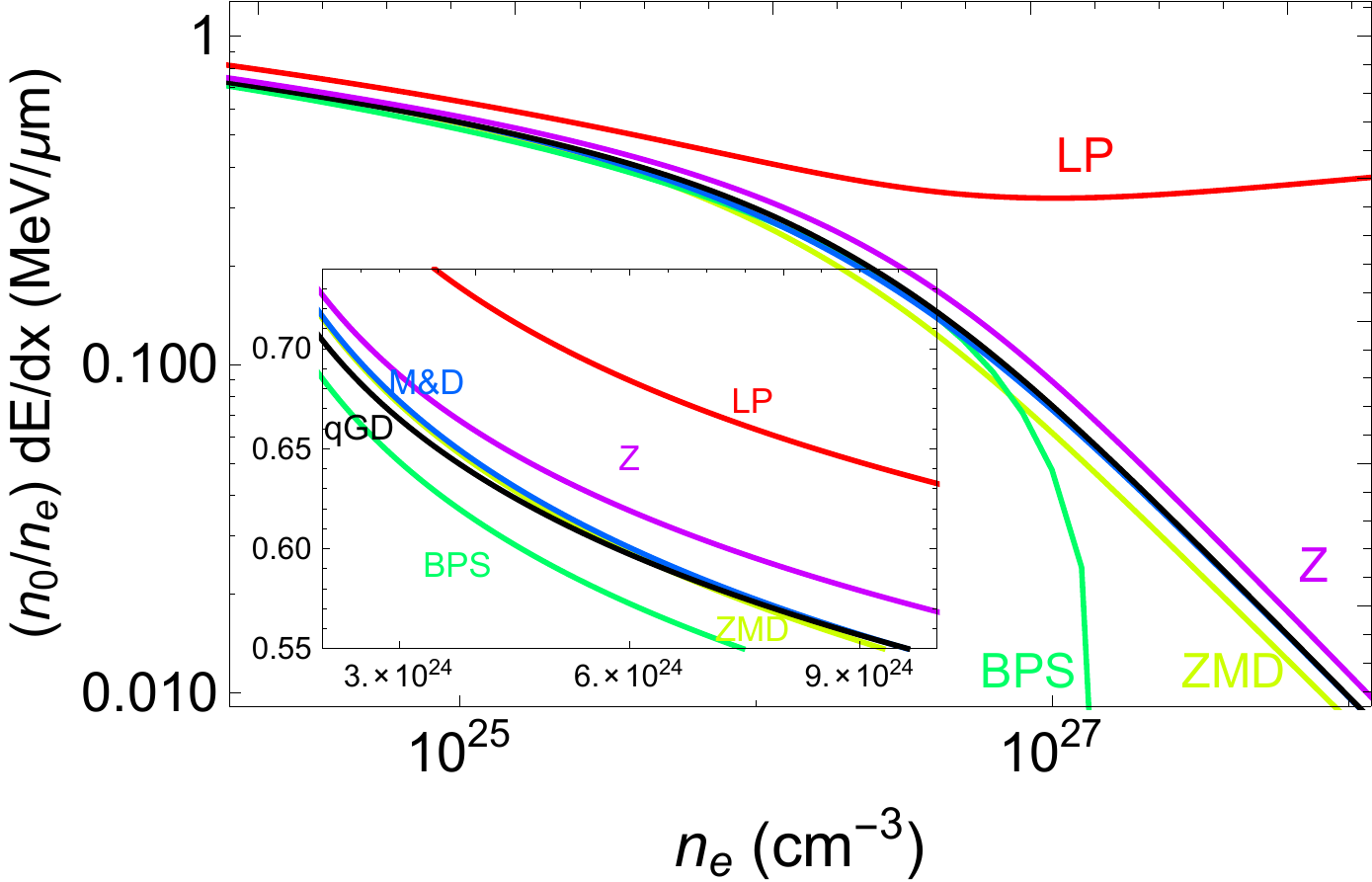}
\caption{\label{NSP} Same as Fig. \ref{ESP} but fixing the energy at $3.5\,$MeV, varying the electron density, and 
scaling the stopping power by $n_0/n_e$, where $n_0 = 10^{25}\,$cm$^{-3}$. We see that neither BPS nor LP have the correct
high density limit since they do not model electron degeneracy.}
\end{figure}

\begin{figure}
\centering
\includegraphics[width=.95\linewidth]{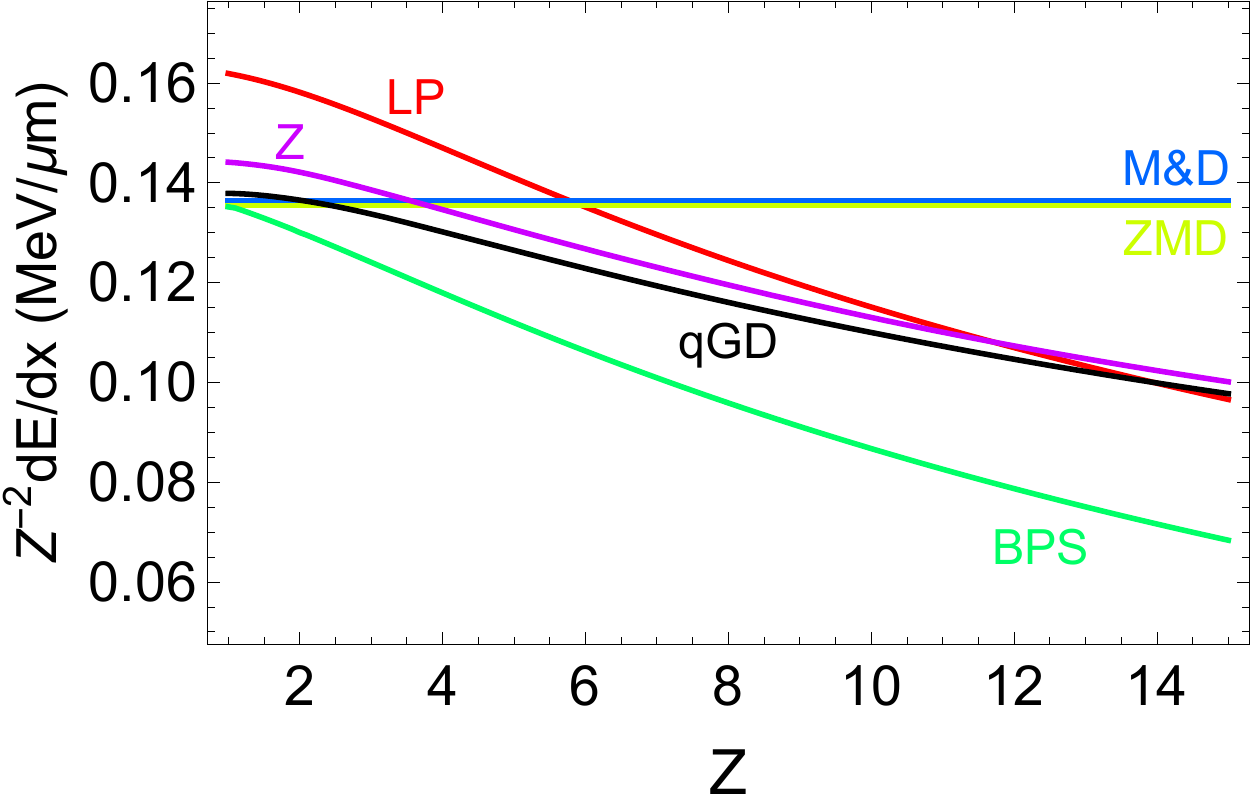}
\caption{\label{ZSP} Same as Fig. \ref{ESP} but fixing the energy at $3.5\,$MeV, varying the alpha particle's charge while keeping
the mass fixed, and 
scaling the stopping power by $Z^{-2}$. Linear models like M\&D and ZMD are unable to capture the strong scattering physics
that becomes more important at high $Z$.}
\end{figure}

To elucidate the many dependencies in \old{this} \new{the} large parameter space, 
we plot several common stopping power models, varying only one parameter at a time. Note, we have simplified matters by doing this exercise for the 
uniform electron gas (no ions, nor bound electrons), which differs from the full test problems of this workshop. 
How to treat bound electrons is a further complicating feature of warm dense matter and an active area of research which increases the uncertainty of stopping power in that regime. 
For this exercise, we focus on alpha particle stopping in hot dense matter. In particular, the conditions are: 
the projectile charge, $Z=2$, electron number density,  $n_e = 10^{25}\,$cm$^{-3}$, electron temperature, $T=1\,$keV, and projectile energy, $E_p=3.5\,$MeV. 
About this common point, one of these four parameters is varied in Figs. \ref{ESP}-\ref{ZSP}. We show six models:
that of Li and Petrasso (LP) \cite{LiPetrasso1993}, including the relatively recent erratum \cite{LiPetrasso1993Err}, 
that of Brown, Preston, and Singleton (BPS) \cite{BrownEtAl2005}, 
that of Maynard and Deutsch (M\&D) \cite{MaynardDeutsch1985}, 
Zimmerman's fit to that model (ZMD) \cite{Zimmerman1990},
the Zwicknagel model (Z) \cite{Zwicknagel1999},
and the quantum mechanical version of the Gould and DeWitt model (qGD) \cite{GouldDeWitt1967}. 
The LP model is derived from weakly coupled plasma theory and includes higher order terms in the Coulomb logarithm.
However, this Coulomb logarithm is given as an ad hoc expression, only valid at weak coupling.
The BPS model uses dimensional continuation analysis to get the weakly coupled, non-degenerate limit accurate to sub-leading order in $g = Ze^2/(\lambda T)$,
where $\lambda$ is the screening length.
The M\&D model depends on linear response theory (small $Z$) and the random phase approximation (weak coupling), but can handle any degeneracy.
The Z model starts with a T-Matrix description, but artificially accounts for dynamic screening and plasmon excitations by multiplying the screening length by $\sqrt{1+v^2/v_{th}^2}$
in the cross section calculation. This gives the Bohr limit at high energies, but is inaccurate when quantum diffraction is important
(the de Broglie wavelength is greater that the classical distance of closest approach).
The qGD model adds together the M\&D model with a T-matrix (strong scattering) model (the latter
is limited to static screening) and subtracts the statically screened version of M\&D so as to not double count (this model is accurate to at least the same order as BPS, but is valid for all degeneracies),

In Fig. \ref{ESP}, we show the dependence of stopping power on energy.  At low energies, the plasma exerts a drag force proportional to the velocity.  At high energies, the plasma does not have
time to react to the projectile and so the stopping decreases. The Bragg peak, where the projectile velocity is roughly the thermal velocity of the target electrons, is when the stopping is maximized. 
We see the the LP model stands out by having a discontinuity near the Bragg peak due to it adding a correction term to get the high energy limit, which is multiplied by a 
Heaviside function. This behavior causes it to overestimate the stopping around the peak. The other models agree with one another within 6\% in this weakly coupled regime.

Figure \ref{TSP} shows the temperature dependence of the models. Since they are all built around the weakly coupled limit, they agree at high temperatures. At low temperatures,
we enter the high energy regime ($E_p\gg T_e$). The correct behavior is that the cold value (which is finite) should be approached. However, the Z and LP models diverge because 
they incorrectly include the classical distance of closest approach instead of the de Broglie wavelength in this limit. Also, at low temperatures, Zimmerman's fit can differ from M\&D, although not catastrophically so. For example in Fig. \ref{TSP}, the difference
is about 18\%.

We also plot the dependence of stopping on density in Fig. \ref{NSP}. We see that the stopping per target electron decreases with density since screening becomes stronger. The differences amongst
models that include degeneracy and those that do not also becomes apparent (Note, that although the absolute difference is small in the plot, the relative difference is 34\% at $n_e = 10^{27}\,$cm$^{-3}$.

The last parameter to vary is the charge of the projectile, which is shown in Fig. \ref{ZSP}. This distinguishes the linear response models (M\&D and ZMD) from those which account for strong scattering.
We see that this effect is insignificant for alpha particles or other common fusion products.

\begin{figure*}
\centering
\includegraphics[width=.95\linewidth]{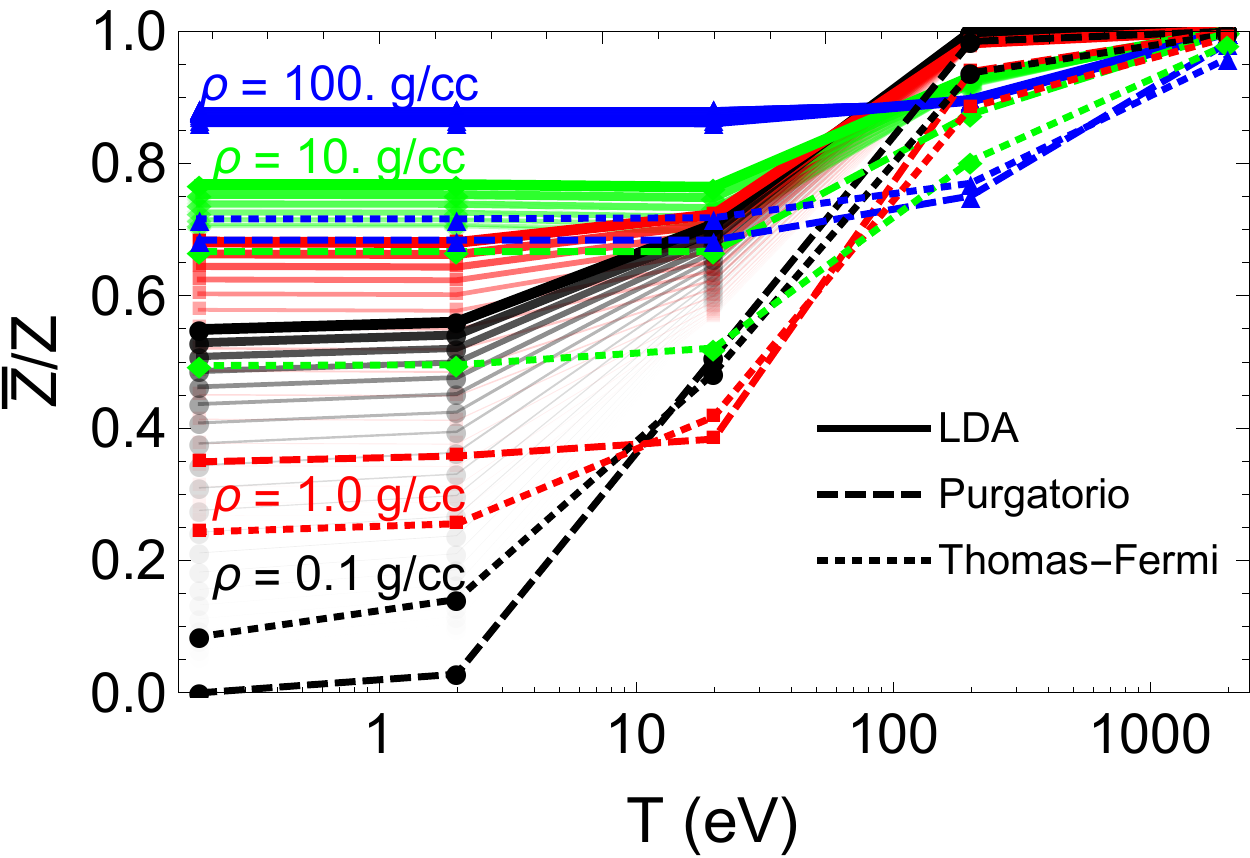}
\caption{\label{Zbar} Different $\bar{Z}$ models for a pure carbon target: the stopping power local density approximation (LDA) using Purgatorio's density and the Zimmerman model for free electrons, 
Purgatorio, an average atom density functional theory code with $\bar{Z}$ calculated from the positive energy electrons, and More's fit to an average atom Thomas-Fermi model. The LDA
prediction is not only a function of density and temperature but also the projectile energy. The uppermost curve of each family is for an alpha particle with $E=3.5$ MeV and the bottommost and
faintest curve is for $E = 500$ eV. Each line is evenly spaced on a logarithmic scale.}
\end{figure*}

All of the above variations are valid for a fully ionized plasma. However, there are often larger uncertainties when there is a 
partially ionized plasma state. There are two common approaches: mixing a bound electron stopping model (usually fit to cold data \cite{SRIM}) and one of the uniform electron gas models above via a value of the mean ionization, $\bar{Z}$, and the local density approximation (LDA).  
Both of these methods require some knowledge of the electronic state, either in $\bar{Z}$ or the electron density. 
Such quantities are acquired via a self-consistent electronic structure calculation  (e.g. density functional theory). 
Zimmerman's model \cite{Zimmerman1990} is an example of the first approach; however, it does not give any opinion on what to use for 
$\bar{Z}$. One option is to use values from More's fit to the electron density at the ion-sphere radius of a Thomas-Fermi atom \cite{More1981} or from a more sophisticated 
equation-of-state code like Purgatorio \cite{Liberman1982,WilsonEtAl2006, SterneEtAl2007}, which outputs the total number of 
electrons with positive energy per atom, but the best practice is to use a $\bar{Z}$ 
designed for the observable of interest. The one approximation used in warm dense matter for stopping power is 
the local density approximation (LDA):
\begin{eqnarray}
\left(\frac{dE}{dx}\right)_{LDA}[n(\mathbf{r})]& =& \frac{Z^2 e^2\omega_p^2}{ v^2} L_{LDA}[n(\mathbf{r})],\\
 L_{LDA}[n(\mathbf{r})]&=&\frac{\int d\mathbf{r}\, n(\mathbf{r}) L_{unif}(n(\mathbf{r})) }{\int d\mathbf{r}\, n(\mathbf{r})},
\end{eqnarray}
where the dependencies on energy and temperature are suppressed for clarity, 
the brackets indicate a functional dependence, the integration is done over a microscopic computational domain
(usually either an average atom or a box containing some tens to hundreds of atoms), 
$\omega_p^2 =  4\pi n_e/m_e$, $n_e$ is the volume-averaged 
electron density, $(dE/dx)_{unif}=Z^2 e^2 \omega_p^2 L_{unif}(n)/v^2$ is the the fully ionized uniform electron gas stopping power (approximations to which were
shown in the Figs. \ref{ESP}-\ref{ZSP}), and $v$ is the projectile velocity. This equation allows us to define a LDA approximation to $\bar{Z}$ via
\begin{equation}
L_{LDA}[n(\mathbf{r})] = L_{unif}\left(\frac{\bar{Z}}{N_I} \sum_i n_i\right),
\end{equation}
where the sum is over all $N_I$ ion species. Here, for computational speed, we use the ZMD model \cite{Zimmerman1990}
of the free electron stopping number for $L_{unif}$ on both sides of the equation. The three predictions of $\bar{Z}$ are 
shown in Fig. \ref{Zbar} for a pure carbon target as a function of temperature for different densities. At low to moderate 
densities ($\lesssim$ 10 g/cc) and low to moderate temperatures ($\lesssim$ 100 eV), the differences between models 
becomes large. Furthermore, the LDA model is dependent on the projectile energy, while the other models
are independent. The accuracy of the LDA model over much of this large parameter space is still largely untested and unknown.

\begin{figure*}
\centering
\includegraphics[width=.95\linewidth]{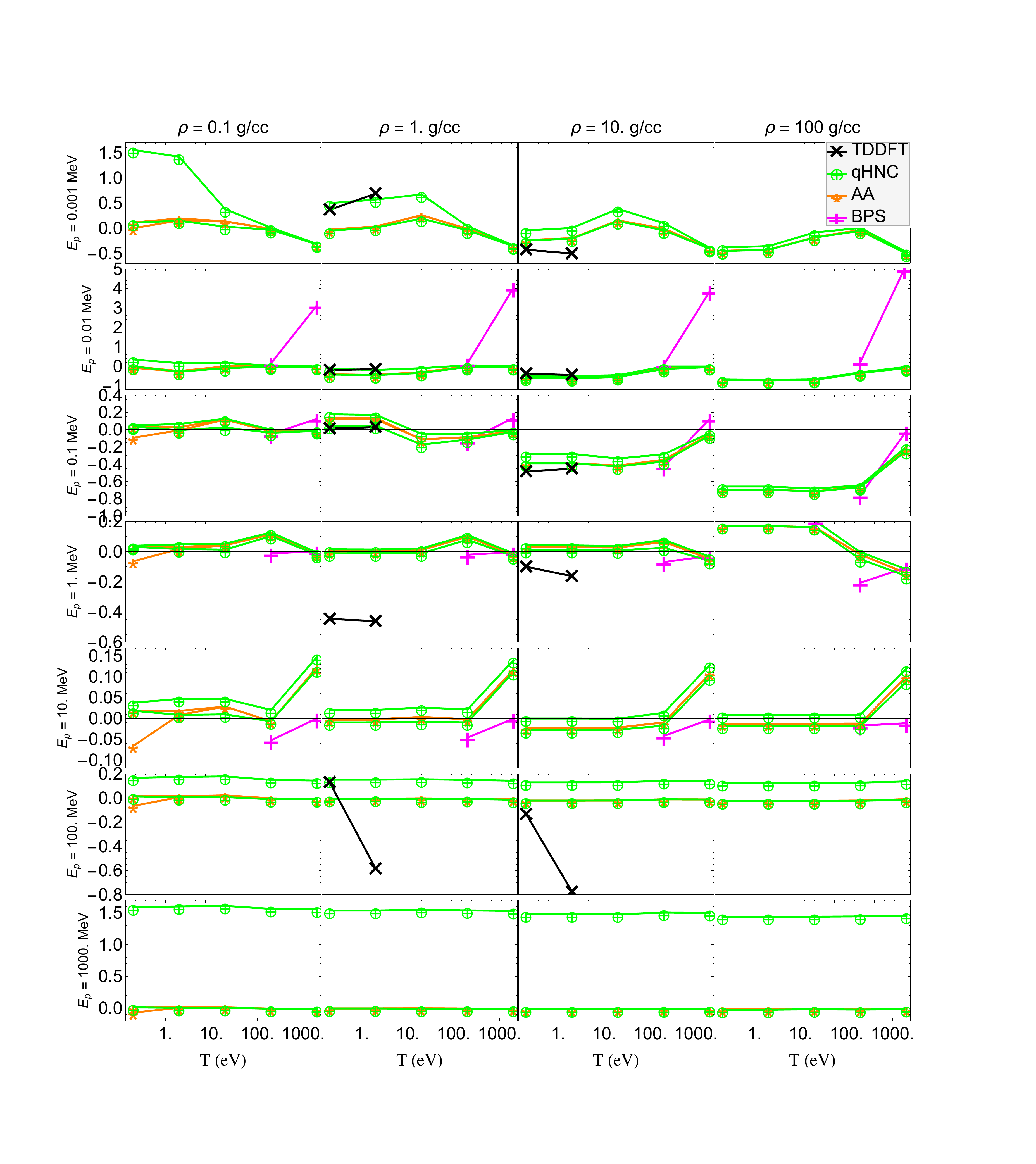}
\caption{\label{SPH} Fractional difference between the electron component of stopping power models and Zimmerman's bound plus free fits \cite{Zimmerman1990}. 
Curves are TDDFT (black), BPS (magenta), 
and various average atom (dashed green) and pseudo-neutral atom (solid green) implementations.
At low energies and high temperatures, the BPS model includes the ion contribution to the stopping, while the 
other models do not. Likewise, the outlier at high energies is actually more accurate since there relativistic and Bremsstrahlung effects become more important.}
\end{figure*}

\begin{figure*}
\centering
\includegraphics[width=.95\linewidth]{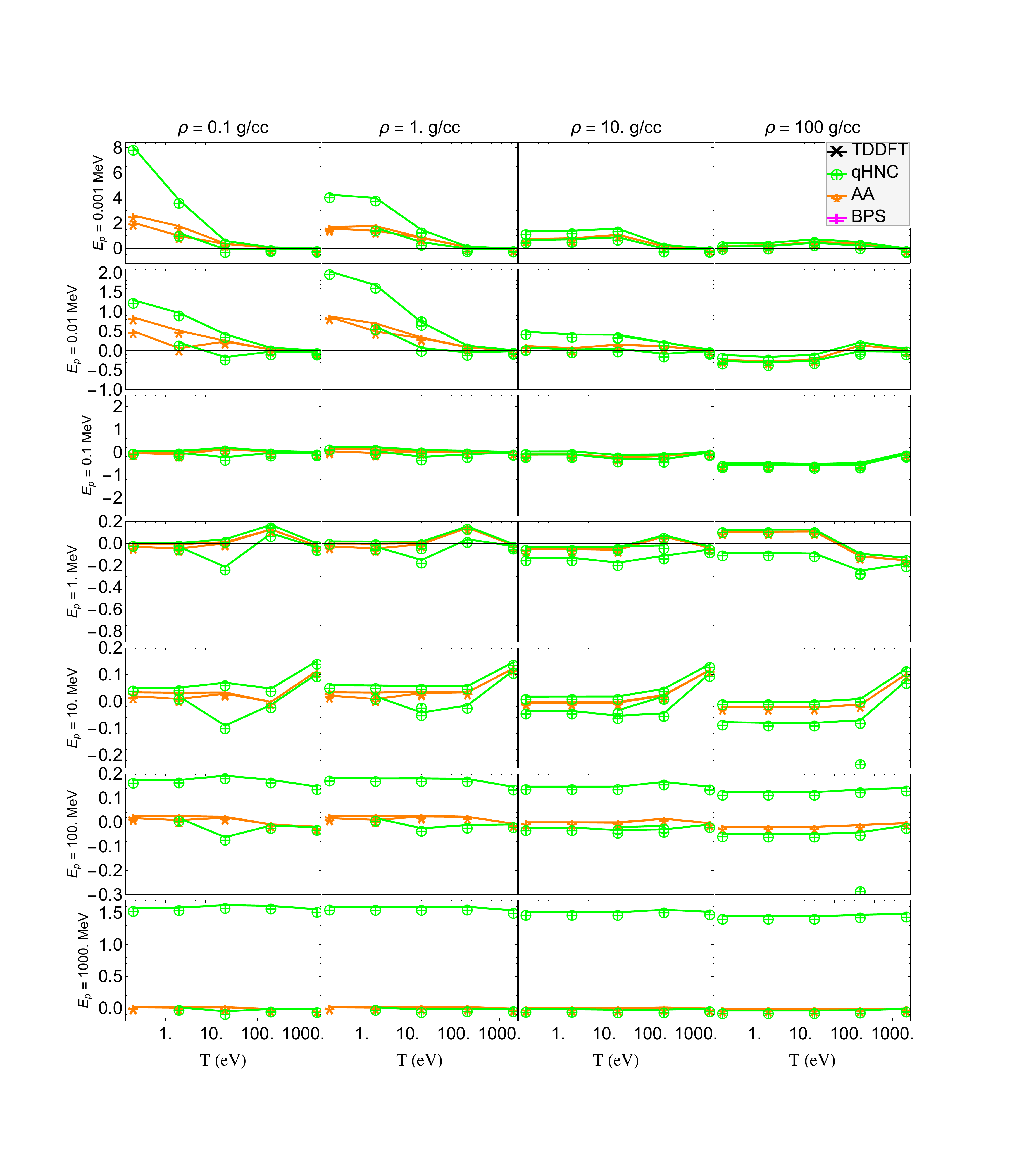}
\caption{\label{SPC} Same as Fig. \ref{SPH} but for carbon.}
\end{figure*}

Finally, we show the results of the code comparison workshop. 
We note that we did not study variations in the ion component of stopping nor different models 
for the charge state of the projectile, both of which can lead to further uncertainties. 
 In order to avoid plotting results that span many orders of magnitude and to emphasize variations amongst the models, we plot 
the relative differences between the submitted results and the full Zimmerman model \cite{Zimmerman1990} (including bound electrons) using More's fit \cite{More1981} to $\bar{Z}$ in Figs. \ref{SPH} and \ref{SPC}
for hydrogen and carbon, respectively (no results were submitted for the mixed CH case). 
Almost all submissions were in the average atom category and proceed to make the same local density approximation to account for the inhomogeneous electron density around each ion. 
So we warn the readers that the variations should be taken as a lower bound for the uncertainty.
The greatest variations are seen at low energy and temperature and high densities. Only the BPS submission attempted to model the ion component of the stopping, which is important at the lower energies
and high temperatures, so for consistency with the majority that part of the Zimmerman model was removed, even though it is
more physical to include it. 
The outlier at high energies is actually more accurate since there relativistic and Bremsstrahlung effects become more important.

\section{Conclusions and Future Workshops}

We have reported on the results of an inaugural charge-particle transport workshop, which aimed to assess variations among model predictions for electronic and ionic transport coefficients and velocity-resolved stopping powers. 
As summarized in Figs. \ref{HAll}-\ref{CHSig}, \ref{SPH}, and \ref{SPC}, we find significant variation (factors of three or more) for all properties among models that represent the range of available theoretical methods, including the values actually used in the hydrodynamic simulations to design and interpret data from ICF experiments. 
Agreement among models was generally higher in the classical weakly coupled regime. At low temperatures and densities, uncertainties in the ionization state were the major source of disagreement. For most transport coefficients, typical best-case variations among codes were 20\% in the weakly coupled regime, factors of two in warm dense matter, 
and worsening to factors of ten or more at low temperatures. 
It is currently not known what the consequences of such uncertainties are, and further integrated modeling of a wide range of experiments is warranted. However, we emphasize that when conduction or diffusion is small, other terms in the hydrodynamic equations will dominate. The result of this workshop may help inform sensitivity studies that would ultimately quantify how important 
these variations are.

It is important to note that this workshop was not intended to identify the ''best'' models. Many of the methods have acknowledged deficiencies or become intractable in some regimes. Rather, the workshop aimed to establish a baseline for model comparison, survey the state-of-the art for a range of \new{modeling approaches and} plasma regimes, provide initial estimates of plausible variations to inform sensitivity studies, and begin to identify the important physics that must be included in reliable transport calculations. Future workshops \old{will explore} \new{would benefit from broader community participation and could examine} in more detail the model assumptions, exploring quantities like interionic and electron-ion potentials, structure factors, and collision cross sections for a more limited set of cases \new{and may eventually expand to include non-equilibrium and high-field effects}. \old{Non-equilibrium and high-field effects may also be of interest in future comparisons. Ideally, these workshops will help establish a foundation for reliable and self-consistent transport calculations which can be used to provide sets of consistent transport and equation-of state properties.} \new{The results from this and any future workshops will help to assess uncertainties in hydrodynamic modeling and establish a foundation for reliable transport calculations. Ultimately, uncertainties and model variations in transport, equation of state \cite{GaffneyEtAl2018}, and radiative properties \cite{nlte} can be rigorously combined to limit as much as possible the potential for inconsistency and offsetting errors.}

\section{Acknowledgements}

\new{We thank Sandia National Labs for hosting the workshop and Steve Haan for providing helpful guidance.
SBH was supported by the U.S. Department of Energy, Office of Science Early Career Research Program, Office of Fusion Energy Sciences under Grant No. FWP-14-017426.}
This work was performed under the auspices of the U.S. Department of Energy by \new{Sandia National Laboratories is managed and operated by NTESS under DOE NNSA contract DE-NA0003525}, Lawrence Livermore National Laboratory under Contract DE-AC52-07NA27344, and Los Alamos National Laboratory under Contract No. 89233218NCA000001. 

\appendix\,

\section{Submitted Data}
For completeness, we are including much of the raw data submitted to the workshop in Tab. \ref{tab:appendix}. Some attendees wished to not make their submissions public, and so those are not included in the table. It is hoped that the data below will help
the community further quantify the range of reasonable values of transport coefficients for a variety of conditions. Brief
descriptions of and references for the methods can be found in Tab. \ref{tab:contributors}. However, we 
encourage researchers to directly contact submitters for details of their calculations.

\onecolumn
\begin{landscape}
\begin{longtable}{|c|c|c|c|c|c|c|c|c|c|c|c|c|}
\caption[Submitted data to the charged particle transport workshop]{\label{tab:appendix}Submitted data to the charged particle transport workshop. 
The columns are the name of the submitter with short descriptor or reference to differentiate multiple submissions from the same person (see Tab. \ref{tab:contributors} for more information), the case written as H for hydrogen, C for carbon, or CH for the carbon-hydrogen mixture followed by two numbers (the first is an index for the density in the set $\{0.1, 1, 10, 100\}$ g/cm$^3$ and the second is an index for the temperature in the set $\{0.2, 2, 20, 200, 2000\}$ eV - for example, C21 refers to the carbon case at 1 g/cm$^3$ and 0.2 eV), electrical conductivity, thermal conductivity, viscosity, diffusion coefficient (self diffusion for the pure substances and inter-diffusion for the mixture), and stopping power at seven different projectile energies. We note that since the 2016 workshop, the values reported here for 
Starrett and Saumon have been superseded by the updated model in Ref \cite{Starrett2017}.} \label{hdata} \\

\hline 
\multicolumn{1}{|c|}{\textbf{Submitter}} & 
\multicolumn{1}{c|}{\textbf{Case}} &
\multicolumn{1}{c|}{\textbf{$\sigma$} }& 
\multicolumn{1}{c|}{\textbf{$\kappa$} }& 
\multicolumn{1}{c|}{\textbf{$\eta$} }& 
\multicolumn{1}{c|}{\textbf{$D$}} &
\multicolumn{7}{c|}{\textbf{$dE/dx$} (\old{MeV}\new{eV}/cm)}\\ 
&& 
 1/($\Omega$ cm) & erg/(s cm K)  & g/(s cm) & cm$^2$/s & 1 keV & 10 keV & 100 keV & 1 MeV & 10 MeV & 100 MeV & 1 GeV \\
\hline 
\endfirsthead

\multicolumn{3}{c}%
{{\bfseries \tablename\ \thetable{} -- continued from previous page}} \\
\hline 
\multicolumn{1}{|c|}{\textbf{Submitter}} & 
\multicolumn{1}{c|}{\textbf{Case}} &
\multicolumn{1}{c|}{\textbf{$\sigma$} }& 
\multicolumn{1}{c|}{\textbf{$\kappa$} }& 
\multicolumn{1}{c|}{\textbf{$\eta$} }& 
\multicolumn{1}{c|}{\textbf{$D$}} &
\multicolumn{7}{c|}{\textbf{$dE/dx$} (\old{MeV}\new{eV}/cm)}\\ 
&& 
1/($\Omega$ cm) & erg/(s cm K)  & g/(s cm) & cm$^2$/s & 1 keV & 10 keV & 100 keV & 1 MeV & 10 MeV & 100 MeV & 1 GeV \\
\hline 
\endhead

\hline \multicolumn{13}{|r|}{{Continued on next page}} \\ \hline
\endfoot

\hline \hline
\endlastfoot

Cl\'erouin (PIJ)                  	&	H11 	&	1.61e3 	&	9.13e5 	&	4.09e-4	&	0.00511	&	      	&	      	&	      	&	       	&	      	&	      	&	      	\\
Hansen (Bemuze)                   	&	H11 	&	       	&	       	&	       	&	       	&	8.37e7	&	4.11e8	&	3.86e8	&	7.33e7 	&	1.06e7	&	1.39e6	&	1.72e5	\\
Hansen (Muze-k)                   	&	H11 	&	1.41e3 	&	7.81e5 	&	       	&	       	&	8.4e7 	&	4.09e8	&	3.85e8	&	7.33e7 	&	1.06e7	&	1.39e6	&	1.72e5	\\
Hansen (Muze-s)                   	&	H11 	&	2.65e-8	&	1.82e-5	&	       	&	       	&	7.54e7	&	3.81e8	&	3.37e8	&	6.65e7 	&	9.72e6	&	1.28e6	&	1.58e5	\\
Haxhimali, Rudd                   	&	H11 	&	       	&	       	&	6.17e-4	&	       	&	      	&	      	&	      	&	       	&	      	&	      	&	      	\\
Copeland \cite{ChapmanCowling1939}	&	H11 	&	       	&	       	&	0.0234 	&	0.14   	&	      	&	      	&	      	&	       	&	      	&	      	&	      	\\
Copeland \cite{PaquetteEtAl1986}  	&	H11 	&	       	&	       	&	1.75e-4	&	0.00285	&	      	&	      	&	      	&	       	&	      	&	      	&	      	\\
Copeland \cite{StantonMurillo2016}	&	H11 	&	       	&	       	&	6.10e-4	&	0.00966	&	      	&	      	&	      	&	       	&	      	&	      	&	      	\\
\end{longtable}
\end{landscape}
\twocolumn



\bibliographystyle{elsarticle-num} 
\bibliography{CPTCW}

%

\end{document}